\documentstyle[epsf]{l-aa}

\def\p{\partial} 
\def\e{{\rm e}}
\def\d{{\rm d}}
\def\ie{i.e. }
\def\eg{e.g. }
\def\etal{et al. }
\def\br{\hat r}
\def\bvr{\bar r}
\def\brs{{\hat r}_{\rm sh}}
\def\ba{\hat a}
\def\M{{\cal M}}
\def\rs{r_\star}
\newlength{\largeur}
\newlength{\saut}
\def\marge#1{
\setlength{\largeur}{\columnwidth}
\addtolength{\largeur}{-#1}
\setlength{\saut}{0.5\largeur}\hspace*{\saut}}
\def\picture #1 by #2 (#3){
 \marge{#1} \vbox to #2{
  \hrule width #1 height 0pt depth 0pt
  \vfill
  \special{picture #3}}}

\begin{document}

\thesaurus{06
(02.01.2;
 02.08.1;
 02.09.1;
 02.19.1;
 08.02.1;
 13.25.5)} 

\title{An analytic study of Bondi--Hoyle--Lyttleton accretion} 
\subtitle{I. Stationary flows} 
\author{T. Foglizzo and M. Ruffert}

\offprints{foglizzo@saclay.cea.fr}

\institute {
Max--Planck--Institut f\"ur Astrophysik, Karl--Schwarzschild--Str. 1, 
Postfach 15 23, 85740 Garching, Germany}

\date{April 26, 1996}

\maketitle
\markboth{An analytic study of Bondi--Hoyle--Lyttleton 
accretion. I}{T. Foglizzo \& M. Ruffert}

\begin{abstract}
We prove that the sonic surface of axi\-symmetric meridional stationary flows
is always attached to the accretor, however small, if
the adiabatic index of the gas is $\gamma=5/3$.\\       
Using local expansions near a point-like accretor, we extend Bondi's
classification of spherically symmetric flows to axisymmetric flows,
introducing the possibility of angular sectors reached by no flow lines, and
singular directions of infinite mass flux, in addition to the angular regions
of subsonic and supersonic accretion. For $\gamma<5/3$, we show the
impossibility of subsonic accretion onto a point--like accretor when the
entropy of the flow is not uniform. The special case $\gamma=5/3$ is treated
separately.\\  
We analyse the influence of the adiabatic index and Mach number of the flow at
infinity on the mass accretion rate of shocked spherical flows. We propose an
interpolation formula for the mass accretion rate of axisymmetric flows as a
function of the Mach number and the adiabatic index, in the range
$9/7<\gamma<5/3$.

\keywords{Accretion, accretion disks -- Hydrodynamics -- Instabilities --
Shock waves -- Binaries: close -- X-rays: stars}

\end{abstract}

\section{Introduction}
Numerical simulations of the Bondi--Hoyle--Lyttleton (hereafter BHL)
accretion flow, in 2--D and 3--D, have not only enabled the determination of
the mean rate of accretion of mass, linear and angular momentum, but have 
also significantly modified our understanding of its dynamics. They have
revealed that the accretion column foreseen by Hoyle \& Lyttleton (1939)
widens to form a detached bow shock if the adiabatic index of the accreted 
gas is $\gamma=5/3$ (Hunt~1971) or $\gamma=4/3$ (Hunt~1979). 
The shock structure observed in simulations was used by Eadie \etal (1975) 
to interpret the X--ray observations of Vel X-1.\\ 
Real situations like wind accretion in X-ray binaries  require taking into
account the density and velocity gradients in the incoming flow (see Ishii
\etal~1993 and references therein), the effects of rotation (see Theuns
\etal~1996 and references therein) as well as the intricacies of radiative
transfer (see an overview in Blondin~1994). For the sake of a better
understanding, we restrict our present investigation to the 
axisymmetric meridional accretion of a gas with uniform adiabatic index onto
a gravitating accretor with constant velocity. \\ 
 Although this configuration is highly simplified, a completely unexpected
non--axisymmetric instability was discovered in numerical simulations for
supersonic flows (Fryxell \& Taam~1988). This instability could explain the
observed variability of the radio or X-ray luminosity in some astrophysical
systems (\eg Ruffert \& Melia~1994).\\  
However, its is not easy to establish the consistency of
the results of the few published 3--D numerical simulations which depend on
the details of the numerical method: no instability appeared in the 3--D SPH
simulations by Boffin (1991) whereas it was observed in the 3--D Eulerian
simulations by Matsuda \etal (1992 and references therein) and Ruffert (1995
and references therein). Dome--shaped structures frequently appeared with
Roe's method (approximate Riemann solver) of Matsuda \etal (1991), but not in
the nested grid Piecewise Parabolic Method (PPM) of Ruffert (1991). Since
numerical viscosity appears to play an important role, a sufficient
resolution of the simulation is crucial. In this respect, numerical
simulations encounter the fo\-llow\-ing difficulty: the smallness of the
accretor in simulations is limited  by the computational power, because the
shortest timestep of the simulation is usually determined by the region of
high velocities in the vicinity of the accretor. The ratio of scales between
the accretion radius and the radius $\rs$ of a compact star, moving with a
supersonic velocity $v$ is typically of order: 
\begin{equation} 
{r_{\rm A}\over \rs}\sim 
10^{5}\left({10^3{\rm kms^{-1}}\over v}\right)^2 \left({r_{\rm Schw.}\over
\rs}\right)\;,\label{ratio5} 
\end{equation} 
where $r_{\rm Schw.}$ is the Schwarzschild radius. Note that in the case of
accretion onto a neutron star, the relevant accretor boundary is the boundary
of its magnetosphere, typically a factor $10-100$ larger than the 
Schwarzschild radius. Some of the most recent 3--D simulations (Ruffert~1992,
1994a) could allow a ratio up to $r_{\rm A}/\rs\sim10^2$ (the full domain of
the simulation extends up to $\sim 16$ accretion radii), by using a PPM code
with multiple nested grids. Such an accreting sphere is still more than 100
times larger than $r_{\rm Schw.}$, and typically $10$ times larger than the
magnetosphere of a neutron star. One would like to be able to predict, on an
analytical basis, which changes are likely to occur if the size of the
accretor is further decreased. The role of the boundary conditions at the
surface of the accretor is closely related to the position of the sonic
surface, and we shall use analytical arguments to obtain some insights on
this question. 

An analytical description of axisymmetric stationary flows is also a first
step towards the analysis of their stability, which we postpone
to a future paper. Bisnovatyi-Kogan \etal (1979) found self-similar
stationary solutions, and one would like to know to what extent any generic
stationary solution would behave in the same way as their solution. More
generality is gained by studying, in the same spirit as Theuns \& David
(1992) for the spherical case, the first order behaviour of all quantities
near the accretor for an axisymmetric flow, and in particular their departure
from sphericity. 

Some analytic formulae have been proposed to fit the mass accretion
rate observed in simulations (Hunt~1979; Ruffert~1994b), but none of them 
filled the gap between the well studied cases of axisymmetric accretion of
dust (Hoyle \& Lyttleton~1939; Bondi \& Hoyle~1944) on the one hand, and
spherical accretion of a gas (Bondi~1952) on the other, for an arbitrary
value of the adiabatic index . Continuing in this direction, we add to these
reference models the spherical accretion of a shocked gas with uniform
adiabatic index, in order to see the effect of both the kinetic and the
thermal energies, and extend this approach to axisymmetric meridional
stationary flows.\\ 

We first recall in Sect.~\ref{Sgeneral} the general equations gover\-ning a
sta\-tion\-ary flow and form\-ulate the axi\-symmetric problem with a single
equation in cylindrical coordinates. The shape of the sonic surface is
analysed in Sect.~\ref{Ssonic}. We establish some properties of the flow by
performing local expansions in the vicinity of the accretor in
Sect.~\ref{Sexpansions}.  We present in Sect.~\ref{Sspherical} the analytical
solution of the spherical accretion from a gas with constant kinetic and
thermal energies at infinity. An interpolation formula for the mass
accretion rate of axisymmetric meridional flows is proposed in
Sect.~\ref{Sinterp}. We summarize our conclusions in Sect.~\ref{Sconclusion}.

\section{Equations of a stationary flow in a gravitational 
potential \label{Sgeneral}} 

\subsection{General equations}

The continuity equation for a gas of velocity ${\vec v}$ and density $\rho$
is:  
\begin{equation}
{\vec\nabla}\cdot (\rho {\vec v})=0\;.
\label{continuity}
\end{equation}
We restrict our study to the case of a gas of uniform adiabatic index
$1<\gamma<5/3$. The entropy $S$ of a gas with pressure $P$ is defined as: 
\begin{equation}
S\equiv{{\cal R}\over(\gamma-1)\mu}\log{P\over\rho^\gamma}\;,
\label{defentro}
\end{equation}
in which ${\cal R}$ is the gas constant and $\mu$ is the mean
molecular  weight. The entropy is constant along each flow line 
(${\vec v}\cdot{\vec\nabla} S=0$), except
when matter passes through a shock.\\
The temperature $T$ and the square of the sound 
velocity $c^2$ are simply related through a multiplicative constant:
\begin{equation}
c^2\equiv {\gamma P\over \rho} ={\gamma {\cal R} T\over \mu}\;.
\label{gparfait}
\end{equation}
For the sake of simplicity, we shall set the ratio ${\cal R}/\mu=1$
everywhere in what follows. Since the mean molecular weight is constant
throughout our study, this convention is equivalent to changing the units of
temperature and entropy, without loosing any generality.\\
Using Eq.~(\ref{defentro}), the pressure force can be written as: 
\begin{equation}
{{\vec \nabla} P\over \rho}={1\over \gamma-1}{\vec \nabla} c^2- 
T{\vec \nabla} S\;.
\label{PTS}
\end{equation}
The Euler equation for a stationary flow with velocity ${\vec v}$ around an 
accretor of mass $M$ is then: 
\begin{equation}
{\vec \nabla} {v^2\over 2} + ({\vec \nabla}\times {\vec v}) \times {\vec v}
=  {\vec \nabla} {GM\over r} - {1\over \gamma-1}{\vec \nabla} c^2 +   
T{\vec\nabla} S\;,   
\label{Euler}
\end{equation}
in which $r$ is the distance to the centre of the accretor and $G$ is the
gravitational constant.\\ 
The Bernoulli equation is obtained by integrating the
Euler equation along  a flow line. It states that the quantity $B$ 
\begin{equation}
B\equiv {v^2\over 2}+{c^2\over \gamma-1}-{GM\over r}\;,
\label{Bernoulli}
\end{equation}
is constant along each flow line. The three terms of this sum are
respectively the kinetic, the thermal, and the gravitational energies.\\
The vorticity of the flow is simply defined as
\begin{equation}
{\vec w} \equiv {\vec\nabla}\times {\vec v}\;.
\end{equation}
By using the definition of the Bernoulli constant (\ref{Bernoulli})
in the Euler equation (\ref{Euler}), we obtain the following equation for any
stationary flow,  
\begin{equation}
{\vec w}\times {\vec v}=T{\vec\nabla} S-{\vec\nabla} B\;.
\label{fmagique}
\end{equation}
This formula is particularly useful because it expresses the vorticity
in terms of invariant quantities $B$ and $S$.\\
The Bernoulli constant $B$ remains constant along a flow line, even 
through a shock. By contrast, the entropy $S$ increases through a shock.\\
In the classic BHL case of a flow with constant and uniform velocity at
infinity,  the Bernoulli constant is uniformly constant (${\vec\nabla}
B={\vec 0}$), and so does not contribute to the vorticity:
\begin{equation}
{\vec w}\times {\vec v}=T{\vec\nabla} S\;.
\label{vortentro}
\end{equation}
If, in addition, the flow is isentropic at infinity, the only possible source 
of vorticity gradients is the shock front. Note that Eq.~(\ref{vortentro})
implies the conservation of entropy (${\vec v}\cdot{\vec\nabla} S=0$). Thus
the only remaining differential equations are the continuity
Eq.~(\ref{continuity}) and the vorticity Eq.~(\ref{vortentro}).

\subsection{Third--order partial differential system of an axi\-symmetric
stationary flow in cylindrical  coordinates}

We consider the axisymmetric stationary flow of a gas with uniform density 
and constant velocity  ${\vec v}_\infty\parallel {\vec x}$ with respect to 
the accretor. The geometry of the flow lines
is described by a single function, the angle $\beta(r,\theta)$ between the
radial direction and the flow line, defined by: 
\begin{equation}
\tan\beta\equiv{v_\theta\over v_r}\;.\label{defbeta}
\end{equation}
We define by ${\vec n}$ the direction perpendicular to the velocity, in the
plane containing the symmetry axis. Denoting by ${\vec e}_r,{\vec e}_\theta$
the unit vectors in cylindrical coordinates, we can write:
\begin{eqnarray}
{\vec v}&\equiv& -v(\cos\beta {\vec e}_r + \sin\beta {\vec e}_\theta)\;,\\
{\vec n}&\equiv& \sin\beta {\vec e}_r - \cos\beta {\vec e}_\theta\;.
\label{defn}
\end{eqnarray}
When useful throughout the paper, we denote by $\bvr$ the dimensionless
distance normalized to the modified accretion radius $r_{\rm E}$, defined
as:  
\begin{equation}
r_{\rm E}\equiv{GM\over
B_\infty}={GM\over{v_\infty^2\over2}+{c_\infty^2\over\gamma-1}}\;,
\end{equation} 
It coincides with the classic accretion radius $r_{\rm A}$ for
large Mach numbers, when the kinetic energy largely exceeds the thermal
energy:   
\begin{equation}
r_{\rm A}\equiv {2GM\over v_\infty^2}\;. \label{rA}
\end{equation}
We shall denote distances normalized to $r_{\rm E}$ by a bar.\\  
Using Eqs.~(\ref{defentro}) and (\ref{Bernoulli}),
all relevant flow quantities can be rewritten as functions of $S,\M,\beta$:
\begin{eqnarray}  
{\vec v}&=& -\M\left\lbrack{2B_\infty(1+\bvr)\over
\bvr\left(\M^2+{2\over\gamma-1}\right)}
\right\rbrack^{1\over 2}(\cos\beta{\vec e}_r+\sin\beta{\vec e}_\theta)
\;,\label{vM}\\ 
\rho&=&\e^{-S}
\left\lbrack{2B_\infty(1+\bvr)\over
\gamma\bvr\left(\M^2+{2\over\gamma-1}\right)}\right\rbrack^{1\over
\gamma-1} \;,\label{rhoM}\\
c&=&\left\lbrack{2B_\infty(1+\bvr)\over
\bvr\left(\M^2+{2\over\gamma-1}\right)}\right\rbrack^{1\over
2} \;,\label{cM}\\ 
P&=&\e^{-S}
\left\lbrack{2B_\infty(1+\bvr)\over
\gamma\bvr\left(\M^2+{2\over\gamma-1}\right)}
\right\rbrack^{\gamma\over\gamma-1}
\;.\label{PM}
\end{eqnarray}
The azimuthal component of the Euler Eq.~(\ref{Euler}), multiplied
by $r$, is: 
\begin{equation}
v_r{\p rv_\theta\over\p r}+v_\theta{\p v_\theta\over\p\theta} +{1\over\rho}
{\p P\over\p\theta}=0\;.
\label{Eulert}
\end{equation}
We denote by $w$ the third and only non-zero component of the vorticity
vector:  
\begin{equation}
w\equiv{1\over r}{\p rv_\theta\over\p r}-{1\over r}{\p v_r\over \p\theta}\;.
\end{equation}
By combining the vectorial vorticity Eq.~(\ref{vortentro}) and the
scalar continuity Eq.~(\ref{continuity}), we obtain a third order
differential system in terms of $\beta,\M^2,S$ only, with variables
$r,\theta$ (see Appendix~\ref{Apartial3}). This system is hyperbolic in
regions where $\M>1$, and elliptic where $\M<1$. It requires {\it three}
functions $\beta_*(\theta),\M^2_*(\theta),S_*(\theta)$ as boundary conditions
on the surface of the accretor. It should be noted that if the entropy is
uniform, the flow is potential (${\vec v}\equiv {\vec\nabla}\Phi_{\rm v}$),
and the Bernoulli and the continuity equations provide a second order partial
differential equation on $\Phi_{\rm v}(r,\theta)$. In this case, the boundary
conditions  at the surface of the accretor consist of {\it two} functions of
$\theta$,  namely $\Phi_{\rm v}(r=\rs,\theta)$ and its radial derivative. 

\subsection{Further reduction of the differential system\label{SScrocco}} 

We can simplify the system by using the modified stream function
introduced by Crocco (1936), defined as follows:  
\begin{eqnarray}
{\p\Psi\over\p\theta}&\equiv&(2GM)^{-{\gamma+1\over2(\gamma-1)}}
c^{2\over\gamma-1}r^2v_r\sin\theta<0\;,\label{dpsit} \\
{\p\Psi\over\p\log r}&\equiv&-\tan\beta{\p\Psi\over\p\theta}\;.
\label{tanbeta} 
\end{eqnarray} 
Where there is no shock, $\Psi$ is conserved along flow lines
(because ${\vec v}\cdot{\vec\nabla}\Psi=0$). It is also conserved across
shocks, since the derivatives of $\Psi$ have only a finite discontinuity
there. The entropy can be expressed as a function of $\Psi$, and thus becomes
a function of a single variable $S(r,\theta)=S(\Psi(r,\theta))$, defined
once and for all along the shock.\\
Let us index the flow lines with their distance $\varpi$ to the symmetry
axis (\ie their impact parameter) at infinity. Only the gas within
a cylinder of radius $\varpi_0$ is accreted, so that the total mass accretion
rate is ${\dot M}=\pi \varpi_0^2\rho_\infty v_\infty$. 
At infinity, the mass flux inside the cylinder of radius $\varpi$ is
directly proportional to $\Psi(\varpi)$ through the following product:
\begin{equation}
\pi \varpi^2\rho_\infty v_\infty=
\pi\left\lbrack{(2GM)^{\gamma+1\over2}\over\gamma}
\right\rbrack^{1\over\gamma-1}
\Psi(\varpi)\e^{-S_\infty}\;.\label{massflux}
\end{equation}
For the sake of clarity of the equations, let us define:
\begin{equation}
\epsilon_\Psi\equiv \bvr{\p\Psi\over\p \bvr}\;.
\end{equation}
Because the total mass accretion rate is finite, we deduce from
Eq.~(\ref{massflux}) that $\Psi$ is bounded in the vicinity of the accretor,
and $\lim_{r\to0}\epsilon_\Psi=0$.\\ 
Using the definition of $\Psi$ and the Bernoulli equation, we can formally
write the Mach number $\M$ as a function of the gradient of $\Psi$ through 
the following algebraic equation: 
\begin{equation}
\left(r\nabla\Psi\right)^2=
{\M^2\sin^2\theta\over r^{5-3\gamma\over\gamma-1}}
\left\lbrack{1+\bvr\over\M^2+{2\over\gamma-1}}
\right\rbrack^{\gamma+1\over\gamma-1}\;.\label{implicit}
\end{equation}
At a given position ($r,\theta$), the maximum of the right hand side term 
is reached at $\M=1$. This defines a maximum value for the gradient of the 
stream function, \ie for the local mass flux according to
Eq.~(\ref{massflux}). For any value of $\nabla\Psi$ below this maximum,
two Mach numbers are solutions of the Eq.~(\ref{implicit}). As for
spherically symmetric flows classified by Bondi (1952), one solution is
subsonic, while the other is supersonic.\\

Equation~(\ref{implicit}) can be derived with respect to $r$
or $\theta$ to express the partial derivatives of $\M$ with respect to $r$ or
$\theta$ as function of $\M$ and the derivatives of $\Psi$. The vorticity
Eq.~(\ref{vortentro}) can therefore be expressed with derivatives of $\Psi$
only, as follows: 
\begin{eqnarray}
&&\bvr^2{\p^2\Psi\over\p \bvr^2}\left\lbrace1+{\M^2\over1-\M^2}\left(
{\epsilon_\Psi\over r\nabla\Psi}\right)^2
\right\rbrace
\nonumber\\
&+&\bvr{\p\Psi\over\p\bvr}\left\lbrace{3\over2}+
{5-3\gamma\over2(\gamma-1)
(1-\M^2)}\right.\nonumber\\
&&\;\;\;\;\;\;\;\;-{\bvr\over1+\bvr}
\left\lbrack{2+(\gamma-1)\M^2\over2(\gamma-1)(1-\M^2)}
\right\rbrack-{\epsilon_\Psi S'(\Psi)\over\gamma\M^2}\nonumber\\
&&\;\;\;\;\;\;\;\;\left.
+{\M^2\over1-\M^2}{1\over\left(r\nabla\Psi\right)^2}
\left(\epsilon_\Psi^2+2{\p\epsilon_\Psi\over\p\theta}{\p\Psi\over\p\theta}
+\epsilon_\Psi{\p^2\Psi\over\p\theta^2}\right)
\right\rbrace\nonumber\\
&+&{1\over 1-\M^2}\left({\p^2\Psi\over\p\theta^2}
-{1\over\tan\theta}{\p\Psi\over\p\theta}\right)
+\left({\p\Psi\over\p\theta}\right)^2
{S'(\Psi)\over\gamma\M^2}=0\;.\label{diffpsi}
\end{eqnarray}
This equation can be viewed as an explicit partial differential equation
of second order on $\Psi$, with $\M$ implicitly depending on $\nabla\Psi$
through the algebraic Eq.~(\ref{implicit}). In an equivalent way,
Eq.~(\ref{diffpsi}) can also be considered as a second order polynom in
$\M^2$, whose two roots explicitly determine $\M^2$ as a function of $\Psi$
and its derivatives. Inserting one of the two roots into
Eq.~(\ref{implicit}), we obtain an implicit differential equation of second
order containing explicitely only $\Psi$ and its derivatives of first and
second order.\\ 
This compact formulation will allow us to perform in Sect.~\ref{Sexpansions} 
a direct analysis of the behaviour of the solution in the vicinity of $r=0$.

\section{The sonic surface of axisymmetric stationary flows
\label{Ssonic}}

\subsection{Three topologies of axisymmetric accretion\label{SStypes}}

\begin{figure} 
\picture 88.9mm by 70.9mm (topology) 
\caption[]{Three topologies of axisymmetric accretion. The sonic
surface is represented as a dotted line. It touches the accretor at an
angle $\theta_{\rm so}$ with the symmetry axis for the types SF and FSF. For
the type FSF, the solid line represents the shock surface reaching the
accretor at an angle $\theta_{\rm sh}$ with the symmetry axis.}        
\label{topologies}      
\end{figure}  

Numerical simulations of axisymmetric accretion with different Mach numbers,
adiabatic indices and accretor sizes have exhibited three different 
topologies of the sonic surface. These three topologies are illustrated in
Fig.~\ref{topologies}.  \par - Type F (fast accretion): the sonic surface is
detached from the accretor. This occurs for subsonic flows for small enough
accretors (Ruffert~1994b, 1995, 1996).    
\par - Type SF (mixed slow/fast accretion): the sonic surface reaches
the surface of the accretor. It subtends an angle $\theta_{\rm so}$ relative
to  the axis of symmetry at the position of the accretor. This is the
case for subsonic flows with large accretors, and supersonic flows with
$\gamma=5/3$ (Matsuda \etal~1992; Ruffert \& Arnett~1994; Ruffert~1994a).  
\par - Type FSF (mixed fast/slow/fast accretion): both the sonic surface and
the shock surface touch the surface of the accretor. They respectively 
subtend the angles $\theta_{\rm so}$ and $\theta_{\rm sh}$ relative to the
axis of symmetry. This occurs for all supersonic flows when the accretor is
large enough, for all the known supersonic 3--D simulations with $\gamma$
close to one (Ishii \etal~1993; Ruffert~1996). More complicated
topologies were observed by  Hunt (1979) and Ruffert (1995) in simulations of
supersonic flows with $\gamma=4/3$, where the bow shock is linked to the
accretor by a conical shock. These can also be classified as Type FSF, with
the difference that the entropy in the supersonic flow ahead of the attached
shock is not uniform.\\  
The topology of the flow in numerical simulations appears to
depend strongly on the size of the accretor: all the supersonic flows are of
Type FSF if the accretor is big enough, but some become Type SF or F when
the accretor size is decreased. 
In order to spare computing time, which increases with decreasing
accretor size, one would like to
use the largest accretor size leading to a physically relevant accretion
process (cf. Eq.~\ref{ratio5}). Type F flows are numerically easier to
handle, because unphysical effects due to a possible reflection of waves at
the boundary of the too large accretor can be avoided.

\subsection{The shape of the flow tubes at the sonic radius\label{SSr0}}

Let us consider a flow tube of cross section $\Phi$, parametrized along the
tube by the curvilinear variable $l$.  
We denote by $\Delta S$ the entropy that this flow tube gains passing
through a shock, and by $\rho_\infty$ and $c_\infty$ the density and the
sound speed at large distances from the accretor, before the shock. The
density $\rho$ and sound speed $c$ after the shock satisfy the following
equation:    
\begin{equation} 
\e^{-\Delta S}\equiv
\left({c_\infty\over c}\right)^{2\over\gamma-1}{\rho\over\rho_\infty}\;. 
\label{entrop}
\end{equation}
Using the Bernoulli equation (\ref{Bernoulli}) and the entropy equation
(\ref{entrop}), we can write the momentum as a function of the Mach number
$\M$ and the position $r$: 
\begin{equation} 
\rho v={\M\rho_\infty\over c_\infty^{2\over\gamma-1}}\e^{-\Delta
S}\left\lbrack {(\gamma-1) B_\infty\left(1+{1\over \bvr}\right)
\over1+{\gamma-1\over 2}\M^2}\right\rbrack^{\gamma+1\over2(\gamma-1)}\;.
\label{momentum} 
\end{equation} 

\begin{figure} 
\picture 86.7mm by 72.3mm (flowtube) 
\caption[]{The position of the sonic point depends on the shape of the
flow tube: for an axisymmetric flow tube, it is shifted from the spherical
position depending on whether its cross section increases faster or slower
than the square of the distance to the accretor }        \label{convexe}  
\end{figure}  

The continuity equation (\ref{continuity}) states that the mass flux is
conserved along the flow tube: 
\begin{equation}
{\p  \over\p l}(\rho v \Phi)=0\;.
\label{dmoment}
\end{equation}
By substituting Eq.~(\ref{momentum}) into Eq.~(\ref{dmoment}), we obtain,
along a given flow line: 
\begin{equation} 
{\p\log\Phi\over\p\log r}=
{\gamma+1\over 2(\gamma-1)}{1\over1+\bvr}-
{1-\M^2\over2+(\gamma-1)\M^2}{\p\log\M^2\over\p\log r}
\;,
\label{consflux}
\end{equation}
where the derivatives are calculated along a flow line, \ie $\p/\p r\equiv
(\p/\p l)(\p l/\p r)$. The regularity of $\p\M/\p l$ at the sonic point
$\M(\bvr_{\rm so})\equiv1$ implies:   
\begin{equation} 
{\p\log\Phi\over\p \log r}=2+
2{r_0-r_{\rm so}\over r_{\rm E}+ r_{\rm so}}\;.
\label{sonic}
\end{equation}
We have denoted by $r_0$ the radius of the sonic sphere in the spherical
case: 
\begin{equation}
r_0\equiv {5-3\gamma\over 4}
{GM\over c_\infty^2+{\gamma-1\over2}v_\infty^2}\;.
\label{rspherique}
\end{equation}
It is remarkable that the spherical sonic radius is a function of
the total energy $B_\infty$ only, and is independent of the entropy of
the flow.\\ 
The regularity condition (\ref{sonic}) is a constraint on the
shape of the flux tube at the sonic radius. Note that this condition is
valid for any stationary flow, with no hypothesis on its symmetry.
Let us illustrate it for an inflowing axisymmetric flow tube: the spherical
case corresponds to radial flow lines, and thus to a conical flow tube. If 
the sonic radius is smaller than the spherical one, the cross section of the
flow tube increases faster than the square of the distance to the accretor,
thus the azimuthal component $v_\theta(r_{\rm so})$ of the velocity at the
sonic radius must be  oriented towards the axis of symmetry. Conversely, if
the sonic radius is larger than the spherical one, the cross  section of the
flow tube increases more slowly than the square of the  distance to the
accretor, and $v_\theta(r_{\rm so})$ must be oriented away from the axis of
symmetry (see Fig.~\ref{convexe}).

\subsection{A geometrical property of Type F flows\label{SSprop}}

\begin{figure} 
\picture 86.7mm by 72.3mm (property) 
\caption[]{If the sonic surface (dashed line) of the axisymmetric flow is
detached from the accretor, it must intersect the sonic sphere (dotted line)
of the equivalent spherically symmetric flow.}         
\label{propriete}   
\end{figure}  
We prove in Appendix~\ref{Aproof} the following property ${\cal P}$:\\
{\it If the sonic surface is detached from the accretor, it must
intersect at least once the sonic sphere of radius $r_0$ of the corresponding
spherically symmetry flow with same energy}. \\
$r_0$ is defined in Sect.~\ref{SSr0} by Eq.~(\ref{rspherique}). This property
can be understood intuitively from Fig.~\ref{propriete}, by using the
geometrical interpretation  of Eq.~(\ref{sonic}) in Sect.~\ref{SSr0}, in the
simple case where the azimuthal velocity is of constant sign along the
sonic surface in the interval $0<\theta<\pi$. If the azimuthal velocity is
directed away from the axis on one side of the accretor 
(for example, $r_{\rm so}(0)>r_0$), it will be oriented towards the axis on
the other side ($r_{\rm so}(\pi)<r_0$), thus implying by continuity an
intersection of the sonic surface with the sonic sphere. We show in  
Appendix~\ref{Aproof} that this property is true for any axisymmetric flow,
without any hypothesis on the distribution of azimuthal velocities.\\ 
Former studies of the BHL accretion have dealt with two set of length scales,
one associated to the Hoyle-Lyttleton accretion of dust (the accretion radius
$r_{\rm A}$), and the other associated to the spherical accretion (the Bondi
radius $r_{\rm B}$ and the corresponding sonic radius $r_0(\M_\infty=0)$).
None of these scales took into account both the kinetic energy and the
thermal energy of the gas. In particular, the significance of the sonic
radius $r_0(\M_\infty=0)$ for axisymmetric flows has always been dubious. The
property ${\cal P}$ establishes the fundamental meaning of the spherical
radius $r_0(\M_\infty)$, in relation with the sonic surface, for axisymmetric
flows of Type F.\\ 
The geometrical property ${\cal P}$ is useful for numerical
simulations of the BHL flow,  in order to fix the maximum size of the totally
absorbing accretor. As discussed in Sect.~\ref{SStypes}, since the sonic
surface must cross the sonic sphere, the accretor size $\rs$ should at least
be smaller than the spherical sonic radius $r_0$. However, although the
sonic sphere is larger than the accretor size used in numerical simulations
with $\gamma=4/3$ and $\gamma=1.01$ (see Ruffert...), the sonic surface and
the shock surface appear to be attached to the accretor. Although necessary,
the condition $\rs<r_0$ is far from being  sufficient.\\

\subsubsection{Consequence for flows with $\gamma=5/3$}

Since $r_0(\gamma=5/3)=0$, the sonic surface of flows with $\gamma=5/3$ must
always reach the surface of the accretor. Following the nomenclature
introduced in Sect.~\ref{SStypes}, {\it the topology of the accretion with
$\gamma=5/3$ is either of Type SF or FSF, for any accretor size and any Mach
number at infinity}.    

\subsubsection{Consequence for nearly isothermal supersonic flows}

The Hoyle-Lyttleton limit for the mass accretion rate at high Mach numbers
requires that the accreted matter comes from a cylinder with a radius scaling
like the accretion radius (Hoyle \& Lyttleton~1939). If $\gamma$ is close to
unity, the sonic spherical radius $r_0$ is $\M_\infty^2/4$ larger than the 
accretion radius $r_{\rm A}$:   
\begin{eqnarray}
{r_0\over r_{\rm A}}&=&{5-3\gamma\over 8}
{\M_\infty^2\over 1+{\gamma-1\over2}\M_\infty^2}\;, \nonumber\\
&\sim& {\M_\infty^2\over 4}\;\;
{\rm for}\;\;\gamma=1\;.
\label{discrep}
\end{eqnarray}
If such a flow were of Type F, the geometrical property ${\cal P}$ would
imply that the sonic surface goes out as far as $\M_\infty^2/4$ times the
accretion radius. This leaves only two possibilities for a stationary flow:
\par(i) the accretor is able to keep material bound up to distances
comparable to $r_0\gg r_{\rm A}$, and the mass accretion rate is infinitely
larger than the Hoyle-Lyttleton limit. This is the case of the spherical
flows we shall study in Sect.~\ref{Sspherical}.
\par(ii) the accretor cannot retain material at such large distances, and the
flow is of Type SF or FSF, as observed in numerical simulations with
$\gamma$ close to one by Ishii \etal (1993) and Ruffert (1996).\\
Let us note that the artificially constructed Type F geometry, where the
distance of the stagnation point exceeds $r_0$, while the accreted matter
collects within a cylinder of radius $\sim r_{\rm A}$, would be strongly
unstable to the Kelvin-Helmholtz instability: the growth rate would
scale like $\sim c_\infty/r_{\rm A}$ while the time needed to reach the
accretor would scale like $\sim c_\infty/r_0$.\\

\begin{figure} 
\picture 86.7mm by 72.3mm (9/7) 
\caption[]{Comparison of the accretion radius $r_{\rm A}$ with the sonic
radius $r_0$ depending on the values of $\gamma$ and $\M_\infty$.
The thin solid line indicates the region of parameters where the
kinetic energy is equal to the thermal energy. The value of $r_{\rm
A}/r_0$ is indicated on the dashed and dotted curves}         
\label{r0rA}     
\end{figure}  

Consequently, we rule out the possibility that the sonic surface is detached
in the case of an isothermal stationary flow. According to
Eq.~(\ref{discrep}), the spherical sonic radius is larger than the
accretion radius at high Mach numbers if (see also Fig.~\ref{r0rA})   
\begin{equation}
1\le\gamma<{9\over 7}\;.
\end{equation}

\section{Asymptotic expansions near the accretor\label{Sexpansions}}

Spherically symmetric accretion flows with $1<\gamma<5/3$ are separated into
two distinct classes, depending on whether the flow is subsonic or supersonic
at the accretor boundary (Bondi~1952). We learnt from Eq.~(\ref{implicit}) in
Sect.~\ref{SScrocco} that these two branches of solutions also exist for
axisymmetric flows, for a given value of the local mass flux at the surface 
of the accretor. The azimuthal parameter $\theta$, however, introduces an
additional degree of freedom, such that the solution might be subsonic in an
angular sector, and supersonic in an other (\ie for a sonic surface attached
to the accretor). We must also distinguish between angular directions
towards which the flow lines converge (direction of ``accreting" flow lines),
and directions along which no gas is accreted (direction of ``passing" flow
lines). By conducting a local analysis, we aim to clarify the various 
possible configurations. 

\subsection{Definition of regular and singular accretion\label{notations}}

For each flow line (indexed by $\varpi$) reaching the accretor, we denote by
$\theta_0(\varpi)$ the angle between the velocity vector ${\vec v}$ and the
axis of symmetry, when $r\to0$. Our notations
are illustrated in Fig.~\ref{nota}. We define the direction $\theta=0$ along
the axis of symmetry on the downstream side of the accretor, so that the
function $\theta_0(\varpi)$ decreases monotonically in the interval
$[0,\varpi_0]$, with $\theta_0(0)=\pi$ and $\theta_0(\varpi_0)=0$. We define
$\theta_{\rm c}$ as the largest azimuthal angle, at $r\to 0$, among all the
converging flow lines, apart from the axis of symmetry:  
\begin{equation}
\theta_{\rm c}\equiv {\rm Max}\{\theta_0(]0,\varpi_0[)\}\;.
\end{equation}
This angle can be smaller than $\pi$. In particular, 
$\theta_{\rm c}<\theta_{\rm sh}$ for flows of Type FSF. Without loosing any
significant physical generality, we can assume that 
$]0,\theta_{\rm c}[\subset \{\theta_0(]0,\varpi_0[)\}$ in the BHL flow.\\
   
\begin{figure} 
\picture 86.7mm by 35.2mm (nota) 
\caption[]{Illustration of our notations $\theta_{\rm c}, \varpi,
\theta_0(\varpi)$}         \label{nota}      
\end{figure}  

Near the surface of the accretor, let us denote by  $2\pi{\dot
m}_0(\theta)\sin\theta\d\theta$ the local flux of mass between the azimuthal
angles $\theta$ and  $\theta+\d\theta$:      
\begin{eqnarray}
{\dot m}_0(\theta)&\equiv &\lim_{r\to0}-r^2 \rho v_r
\;,\label{locflux}
\\
&=&\lim_{r\to0}
-\left\lbrack{(2GM)^{\gamma+1\over2}\over\gamma}
\right\rbrack^{1\over\gamma-1}
{\e^{-S}\over\sin\theta}{\p\Psi\over\p\theta} \;,\label{mpsi}\\ 
&=&\lim_{r\to0}{\e^{-S}\M\cos\beta\over
\gamma^{1\over\gamma-1}r^{5-3\gamma\over2(\gamma-1)}}
\left\lbrack{2GM(1+\bvr)\over
\M^2+{2\over\gamma-1}}
\right\rbrack^{\gamma+1\over
2(\gamma-1)} \label{mM}\;.
\end{eqnarray}
Since the stream function is bounded in the flow (Eq.~\ref{massflux}),
integrating Eq.~(\ref{tanbeta}) with respect to $r$ implies:   
\begin{equation}
\lim_{\rs\to0} \beta{\p\Psi\over\p\theta}=0\;\;{\rm
for}\;\theta\in[0,\pi]\;. \label{vrvt}
\end{equation}
Thus, if the mass accretion rate ${\dot m}_0$ is locally not zero, the flow 
lines become locally radial when $r\to0$ (no ``spiraling" of the flow lines).
On the other hand, if the flow lines are not radial, Eq.~(\ref{mpsi}) implies
that the mass accretion rate is zero in this direction (passing flow 
lines).\\ 
Note that along a direction of accretion, we can consider a flux
tube of cross section $\Phi_\varpi$ around the flow line such that
$\theta_0(\varpi)=\theta$, and write Eq.~(\ref{locflux}) as follows:  
\begin{equation}
{\dot m}_0(\theta_0(\varpi))= -(\rho v\Phi_\varpi)\; \lim_{r\to0} 
{r^2\over \Phi_\varpi}\;.\label{regular}
\end{equation}
The quantity ${\dot m}_0$ is consequently finite only if the flow tubes are
asymptotically conical ($\Phi_\varpi(r)\sim r^2$). This leads us to call 
{\it regular} the accretion onto a point--like accretor such that  ${\dot
m}_0$ remains finite, and {\it singular} when it is locally infinite.\\
We can also express the local mass accretion rate in terms of the mass flux
between the cylinders of radii $\varpi$ and $\varpi+\d\varpi$ at infinity:
\begin{equation}
{\dot m}_0(\theta_0(\varpi))=
-{\rho_\infty v_\infty\over{{\sin\theta_0\over\varpi}
{\d\theta_0\over\d\varpi}} }\;.\label{regular2}
\end{equation}
Thus singular accretion corresponds to a localized azimuthal
angle $\theta_0$ where $\d\theta_0/\d\varpi=0$. An example of singular
accretion is the self-similar solution found by  Bisnovatyi-Kogan \etal
(1979), where the whole accretion is along a single azimuthal angle
$\theta_{\rm k}=\theta_0(]0,\varpi_0])$.\\ 
For a regular flow ($\Phi_\varpi(r)\sim
r^2$), we take the limit of Eq.~(\ref{consflux}) when $r\to0$ and obtain,
along a given flow line,
\begin{equation}
{\p\log\M^2\over\p\log r}\sim {5-3\gamma\over2(\gamma-1)}
{2+(\gamma-1)\M^2\over 1-\M^2}\;.
\end{equation}
This equation shows that for $\gamma<5/3$, the Mach number along regular flow
lines either tends to zero or to infinity when approaching the accretor. As 
in the case of spherical accretion (Bondi~1952) no intermediate finite Mach
number is possible. Moreover, this equation proves, by
continuity, that a mixed subsonic/supersonic accretion requires a singular
accretion at the transition if $\gamma<5/3$. The sonic surface may still
reach the accretor in a regular flow, by separating a region of supersonic
accretion from a subsonic region of passing flow lines, as we shall see in
Sect.~\ref{SSinterface}.\\ 
We will now examine the different regions of passing flow lines, supersonic
and subsonic regular accretion successively, for $\gamma<5/3$, and treat the
case $\gamma=5/3$ separately. We shall use the generic name $\epsilon(r)$ for
a function which tends to zero with $r$, and the subscript $0$ for the limit
of functions when $r\to 0$.  

\subsection{Region of passing flow lines $\theta\in]\theta_{\rm c},\pi[$
for $\gamma\le5/3$\label{noaccretion}}  
Since the flux of mass is zero, $\beta_0(\theta)$ cannot vanish in the
interval $]\theta_{\rm c},\pi[$. According to our sign convention in
Sect.~4.3.3 and the definition of $\beta$ in Eq.~(\ref{defbeta}), 
\begin{equation} 
\beta_0(\theta)>0\;\;{\rm for }\;\theta\in]\theta_{\rm
c},\pi[\;. 
\end{equation}
Since $\beta_0(\pi)=\beta(\theta_{\rm c})=0$, the function $\beta_0$ must
have a maximum in the interval $]\theta_{\rm c},\pi[$.\\
\par (i) If the shock is detached, the entropy is finite in the flow.
According to Eq.~(\ref{mM}), the Mach number must either diverge or tend to
zero in order to satisfy ${\dot m}_0=0$.\\
Let us suppose that $\M\to\infty$. We use Eq.~(\ref{vM}) and obtain: 
\begin{equation}
w=-{v\over 2r}\sin\beta_0\left(1+2{\p\beta_0\over\p\theta}\right)
(1+\epsilon(r))\;.
\end{equation}
Using the vorticity equation (\ref{vortentro}), the entropy gradient can be
expressed as:
\begin{equation}
\nabla S={\gamma\M^2\over2r}
\sin\beta_0\left(1+2{\p\beta_0\over\p\theta}\right)(1+\epsilon(r))\;.
\end{equation}
Integrating this equation over a segment between $\theta=\pi$ and 
$\theta=\theta_{\rm max}$, for $r\to0$, contradicts the fact that the
entropy is bounded for a flow with a detached shock. Consequently the Mach
number tends to 0 when $r\to 0$, and 
$\theta_{\rm so}\le \theta_c$. {\it We conclude that all the flows with a
detached shock and $\theta_{\rm c}<\pi$ are of Type SF.}\\
\par(ii) If the shock is attached, the incoming supersonic velocity diverges
near the accretor like $r^{-1/2}$ when $r\to0$. 
We decompose it into a parallel and a perpendicular
component with respect to the shock. The Mach number $\M_{2\perp}$ associated
to the component of the velocity, after and perpendicular to the shock, is
subsonic according to the Rankine--Hugoniot jump conditions (see \eg Landau
\& Lifschitz~1987):   
\begin{equation}
\lim_{r\to0} \M_{2\perp}(\theta_{\rm sh})={\gamma-1\over2\gamma}\;.
\end{equation}
The Mach number $\M_{2\parallel}$ associated to the component parallel to the
shock can be supersonic, but not arbitrarily large:
\begin{equation}
\lim_{r\to0} \M_{2\parallel}(\theta_{\rm sh})=
{\gamma+1\over\left\lbrack2\gamma(\gamma-1)\right\rbrack^{1\over2}} 
{1\over\tan\beta(\theta_{\rm sh})}\;.
\end{equation}
{\it The total Mach number immediately after the attached shock is
consequently finite when $r\to0$}. This result will be important in order to
study the local stability of such flows to the Rayleigh--Taylor instability.

\subsection{Region of regular supersonic accretion for $\gamma<5/3$
\label{SSsuper}}   

If $\M\to\infty$ and ${\dot m}_0$ is finite,
Eqs.~(\ref{vM})--(\ref{PM}) can be expanded as follows:   
\begin{eqnarray} 
v_r&=& -\left({2GM\over r}\right)^{1\over 2}(1+\epsilon(r))
\;,\label{vrsup}\\
\rho&=&{\dot m}_0
\left({1\over 2GMr^3}\right)^{1\over2}(1+\epsilon(r)) 
\;,\\
c&=&\left(\e^{S_0}{\dot m}_0\right)^{\gamma-1\over2}
{\gamma^{1\over2}\over(2GMr^3)^{\gamma-1\over4}} (1+\epsilon(r)) \;,\\ 
P&=&\e^{(\gamma-1)S_0}{\dot m}_0^\gamma
 \left({1\over2GMr^3}\right)^{\gamma\over2}(1+\epsilon(r))
\;,\label{Psup}\\ 
\M&=&\left({\e^{-S_0}\over
{\dot m}_0}\right)^{\gamma-1\over2}
{(2GM)^{\gamma+1\over4}\over\gamma^{1\over2}
r^{5-3\gamma\over4}}(1+\epsilon(r))
\nonumber\\ 
&&\to \infty\;{\rm when}\;r\to 0\;. 
\end{eqnarray}
The first order azimuthal velocity can be estimated
using Eq.~(\ref{vrvt}) in the azimuthal Euler Eq.~(\ref{Eulert}). If the
pressure is not spherically symmetric to first order,   
\begin{eqnarray}
{1\over\rho}{\p P\over \p\theta}&=& -{3\over 2}(2-\gamma) v_r
v_\theta(1+\epsilon(r))\;, \label{prevt}\\
v_\theta &=&{\cal O}\left(r^{4-3\gamma\over2}\right)\;.\label{vtheta} 
\end{eqnarray}
$v_r$ being negative, the azimuthal velocity is oriented towards the region
of increasing pressure. The azimuthal pressure force is balanced by this
dynamical term. We deduce from Eqs.~(\ref{defbeta}) and (\ref{vtheta})
an upper bound for the departure from spherical
symmetry of the function $\beta$:
\begin{equation}
\lim_{r\to0}\left({\p\log\beta\over\p\log r}\right)\ge{5-3\gamma\over2}\;.
\label{betamax}
\end{equation} 
By studying the leading terms of Eq.~(\ref{diffpsi}) when $r\to0$, we derive
in Appendix~\ref{Ainf53} a lower bound for the departure from spherical
symmetry of the function $\beta$, if the entropy of the flow is not uniform:
\begin{equation}
\lim_{r\to0}\left({\p\log\beta\over\p\log r}\right)\le 5-3\gamma\;
\;\;{\rm if }\;S'(\Psi)\ne0\;.
\label{betamin}
\end{equation} 
On the other hand, the function $\beta$ may approach 0 much more rapidly
than $\sim r^{5-3\gamma}$ if the entropy is uniform. To phrase it more
qualitatively, the presence of an entropy gradient sets a lower bound on the
``bending" of the flow lines near the accretor. If a parallel can be drawn
with the simpler case of two-dimensional supersonic flow past a concave
profile (\eg Landau \& Lifshitz 1987), this ``bending" could be related to
the formation  of long--lived radial shocks in the supersonic region of flows
with a non--uniform entropy, as observed in numerical simulations 
with $\gamma=4/3$ in 2--D (Hunt 1979) and 3--D (Ruffert~1995).\\

Because the radial velocity (\ref{vrsup}) is spherically symmetric to first
order, the azimuthal dependence of the accreted momentum $\rho v$ comes
mainly from the azimuthal dependence of the density $\rho$, which we
decompose into  the contributions of pressure and entropy, 
\begin{equation}
\rho(r,\theta)=P^{1\over\gamma}(r,\theta)\exp\left(-{\gamma-1\over\gamma}
S(r,\theta)\right)\;.
\label{densentr}
\end{equation}
Thus we expect that for a flow with uniform entropy (no shock), the mass is
mainly accreted from the hemisphere where the pressure is the highest.
The geometrical interpretation of Eq.~(\ref{sonic}) implies that near the
axis of symmetry (\ie for $\theta\sim0$ and $\theta\sim\pi$), the azimuthal
speed is oriented away from the axis where $r_{\rm so}>r_0$, and towards the
axis where $r_{\rm so}<r_0$. If one assumes that the sign of the azimuthal
velocity does not change in the supersonic region between the sonic surface
and the accretor, Eq.~(\ref{prevt}) indicates that the highest pressure is on
the side of the accretor where $r_{\rm so}>r_0$.\\ 
For an isentropic subsonic flow, the natural orientation
of the azimuthal velocity is set by the sign of the angular momentum of the
fluid element at infinity. A subsonic medium
would flow from the {\it left} of the accretor in Fig.~\ref{propriete}.
Therefore the sonic surface is displaced opposite to the direction of the
flow at infinity, and the maximum flux of mass is expected on the back
hemisphere of the accretor. Indeed, in the subsonic numerical simulations of
a flow with $\gamma=4/3$ by Hunt (1979) in 2--D, and by Ruffert (1995) in
3--D, the sign of the accreted linear momentum is negative for the smallest
accretor considered.\\  
If the flow is supersonic at infinity, the position of the shock along the
axis is displaced in the direction of the flow, compared to its position for 
a spherical flow with same kinetic and thermal energies at infinity, because
there is now a leakage of gas away from the accretor and the pressure ahead
of the accretor is lower than in the corresponding spherical case. The sonic
point along the axis is displaced accordingly. The sonic surface is
consequently shifted in the direction of the flow, leading to a highest
pressure on the front hemisphere of the accretor. The simplest configuration 
is the one in Fig.~\ref{propriete}, where the supersonic flow now
comes from the {\it right} of the accretor.\\ 
Nevertheless, the entropy generated
by the shock decreases with increasing distance from the axis of symmetry
(see Appendix~\ref{Aaxis}), so that entropy and pressure contribute in
opposite directions to the density in Eq.~(\ref{densentr}), precluding a
direct conclusion about its azimuthal distribution near the accretor.
However, the accreted linear momentum is  negative in the numerical
simulations by Ruffert (1996 and references therein) indicating that  
the entropy effect is stronger than the pressure effect.

\subsection{Region of regular subsonic accretion for $\gamma<5/3$
\label{Ssubsonic}} 
If $\M\to0$ and ${\dot m}_0$ is finite,
Eqs.~(\ref{vM})--(\ref{PM}) can be written as follows:      
\begin{eqnarray}
v_r&=& \e^{S_0}{\dot m}_0
\left\lbrack{\gamma\over(\gamma-1)GM}\right\rbrack^{1\over \gamma-1} 
r^{3-2\gamma\over\gamma-1}(1+\epsilon(r))
\;,\label{vrsub}\\
\rho&=&\e^{-S_0}
\left\lbrack{(\gamma-1)GM\over \gamma r}\right\rbrack^{1\over\gamma-1} 
(1+\epsilon(r))
\;,\label{rhosub}\\
c&=&\left\lbrack{(\gamma-1)GM\over r}\right\rbrack^{1\over2}(1+\epsilon(r))
\;,\label{csub}\\
P&=&\e^{-S_0}
\left\lbrack{(\gamma-1)GM\over \gamma r}\right\rbrack^{\gamma\over \gamma-1}
(1+\epsilon(r)) 
\;,\label{Psub}\\
\M&=& 
{\gamma^{1\over \gamma-1}\e^{S_0}{\dot m}_0
\over\left\lbrack(\gamma-1)GM
\right\rbrack^{\gamma+1\over \gamma-1}}
r^{5-3\gamma\over 2(\gamma-1)}(1+\epsilon(r))
\nonumber\\
&&\to 0\;{\rm when}\;r\to 0\;. \label{machsub}
\end{eqnarray}
We use Eqs.~(\ref{vrsub}) and~(\ref{vrvt}) to estimate the dynamical
terms balancing the azimuthal pressure force in Eq.~(\ref{Eulert}) and
obtain: 
\begin{equation} 
{1\over\rho}{\p P\over\p\theta}\ll{\cal O}
\left(r^{2(3-2\gamma)\over\gamma-1}\right)\ll{\cal O}
\left(r^{-1}\right)\;. 
\label{maxiP}
\end{equation} 
Comparing this to the first order of $P/\rho\sim c^2$ given
by Eq.~(\ref{csub}), we conclude for $\gamma<5/3$ that {\it the pressure in
the  regular subsonic region is spherically symmetric to first order when
$r\to 0$}.\\ 
Thus Eq.~(\ref{Psub}) implies that the subsonic flow for $\gamma<5/3$, with
its first order expansions (\ref{vrsub})--(\ref{machsub}),
is possible {\it only if the entropy is uniform}. \\
In particular, the spherical solution proposed by Theuns \& David (1992),
connecting a supersonic solution to a subsonic solution through a shock,
cannot resist the slightest asphericity if the accretor is point--like.
The azimuthal pressure force due to the convergence of flow lines of 
different entropies is too big to be balanced by dynamical forces in the
regular subsonic region close to the accretor. Nevertheless, this constraint
of the azimuthal Euler equation is less stringent for large accretor sizes.
One can also argue that the radiative processes near the surface of the star
may homogenize the entropy, and allow for a spherically symmetric subsonic
accretion. \\
The entropy in flows of Types SF and FSF is a {\it strictly} decreasing
quantity along the shock:
\par(i) In Type FSF flows, the velocity diverges to infinity when $r_{\rm
sh}\to0$, thus the entropy created along  the shock decreases strictly with
distance from the accretor for Type FSF flows, precluding a subsonic
accretion with a uniform entropy.  
\par(ii) For Type SF flows, we show in Appendix~\ref{Aaxis} that the entropy
created along the shock strictly decreases with distance from the axis of
symmetry.\\ 
We conclude that there cannot be a region of regular subsonic accretion in
shocked flows with $\gamma<5/3$, for a point--like accretor. The subsonic
region may reach the surface of the accretor for any finite size, but
{\it the mass accreted in the shocked regular subsonic region
$\theta>\theta_{\rm so}$ tends to zero when the accretor size decreases to
zero}. 

\subsection{The sonic surface for Type SF and FSF shocked flows
$\gamma<5/3$ \label{SSinterface}} 
Since no regular subsonic accretion is possible with a non--uniform entropy,
all the regular flow lines are ultimately supersonic when they reach the
point--like accretor $(\theta_{\rm c}\le\theta_{\rm so})$. Note that in the
singular solutions by Bisnovatyi-Kogan \etal (1979), the  Mach number tends 
to infinity along each flow line.\\ 
Together with the results of Sect.~\ref{noaccretion}, we conclude that for
a flow with a detached shock, and $\gamma<5/3$,
\begin{equation}
\theta_{\rm c}=\theta_{\rm so}\;.
\end{equation}
According to Eq.~(\ref{sonic}), the cross section of the
flow tube reaching the sound speed at the surface of the accretor varies
like:      
\begin{equation}
{\p\log\Phi\over\p l}{\p l\over\p\log r}=
{\gamma+1\over 2(\gamma-1)}>2\;.
\label{dphi}
\end{equation}
Comparing this equation to Eq.~(\ref{regular}), we conclude that the
azimuthal angle $\theta_{\rm so}$ cannot be a direction of regular
accretion. However, $\theta_{\rm so}$ is not necessarily a direction
of singular accretion. A regular flow simply requires that $\theta_{\rm
c}\notin \{\theta_0(]0,\varpi_0[)\}$.

\subsection{The particular case $\gamma=5/3$ \label{S53}}

We denote by $\M_0(\theta)$ the limit of the Mach number along a converging
flow line. For $\theta\in]0,\theta_{\rm c}[$, Eq.~(\ref{mM}) implies that the
local mass accretion rate ${\dot m}_0$ is always finite, its maximum value 
being reached for $\M_0=1$: 
\begin{eqnarray}
{\dot m}_0&=&\left({3\over5}\right)^{3\over2}
4 G^2M^2\e^{-S_0}
{\M_0\over\left(\M_0^2+3\right)^2}\;,\label{m53} \\
&\le&\left({3\over5}\right)^{3\over2}{G^2M^2\over4}\e^{-S_0}\;.
\end{eqnarray}
The pressure forces for $\gamma=5/3$ are strong enough to impede an
arbitrarily high mass flux, even locally. Thus accretion flows with
$\gamma=5/3$ are always regular, and the maximum mass accretion rate is the
spherical Bondi value. This  result is particularly important to
understand the instability observed in numerical simulations. Since
the most unstable simulations were observed with $\gamma=5/3$, we can
restrict our future stability analysis to the simpler case of
regular flows.\\ 
Using Eq.~(\ref{vM}) and~(\ref{vrvt}) in the azimuthal
Euler Eq.~(\ref{Eulert}), we find that the azimuthal pressure force is
bounded by Eq.~(\ref{maxiP}). Comparing it to Eqs.~(\ref{rhoM})
and~(\ref{PM}), we conclude that the pressure is spherically symmetric to
first order for $\theta\in]0,\theta_{\rm c}[$. According to
Eq.~(\ref{densentr}), the density is spherically symmetric to first order if
there is no shock (uniform entropy), but increases towards the back
hemisphere if there is a shock.\\  
With Eqs.~(\ref{vM}) to~(\ref{PM}), we conclude that: 
\par(i) in the absence of a shock, the pressure, density, temperature,
velocity and Mach number are spherically symmetric to first order when $r\to
0$. The accreted momentum is therefore spherically symmetric as well. This is
confirmed by the numerical simulations by Ruffert (1994b), with
$\M_\infty=0.6$ and $\gamma=5/3$. The accreted linear momentum decreases to
zero with the size of the accretor (see Fig.~2, model ``SL", ``SM", ``SS").
Moreover, the  uniform Mach number $\M_0$ must be lower than or equal to one
because of the property ${\cal P}$.
\par (ii) if a shock is present, the pressure is spherically symmetric to
first order when $r\to 0$, the temperature and entropy increase towards the
upstream hemisphere, and the density, the velocity and the Mach number
increase towards the downstream hemisphere. Consequently, more mass is
accreted from the downstream hemisphere than from the upstream hemisphere.
This is also confirmed by the numerical simulations with $\gamma=5/3$ by
Ruffert (1994b), at $\M_\infty=1.4$. This trend persists in the unstable
cases, simulated by Ruffert \& Arnett (1994) at $\M_\infty=3$, and Ruffert
(1994b), at $\M_\infty=10$.\\ 
Thus we disagree with Hunt (1971), who chose the Mach number at the surface
of the accretor $\M_0$ in order to maximize the mass accretion rate. As can 
be checked from Eq.~(\ref{m53}), the highest mass accretion rate is obtained
for $\M_0=1$. The Mach number at the surface of the accretor is, however, not
a free parameter, but is constrained by the distribution of entropy. 
Let $S_1$ be the entropy of the flow line reaching the accretor with
$\M_0=1$. The pressure being uniform,  Eq.~(\ref{PM}) implies that the final
Mach number of the flow line indexed by $\varpi$ depends on its
entropy $S(\varpi)$ according to:       
\begin{equation}
\M_0^2(\varpi)=4\e^{{2\over5}\left(S_1-S(\varpi)\right)}-3 \;.
\label{machaccr}
\end{equation}
The erroneous conclusion of Hunt (1971), that the axisymmetric flow with
$\gamma=5/3$ never becomes supersonic close to the accretor is in
contradiction with the 3--D numerical simulations (see Fig.~17 in 
Ruffert~1994b). It is even in contradiction with the Mach number 
$\M_0(\theta=\pi)\sim 2.1$ that one can deduce from the velocity, entropy 
and density displayed in Figs.~5, 6 and 8 in Hunt (1971) for $\M_\infty=2.4$.
\\  
In a similar way as Eq.~(\ref{SbMlies}), obtained for $\gamma<5/3$,
Eq.~(\ref{machaccr}) shows that the functions $S,\beta,\M$ also cease to be
independent at the boundary of the accretor for $\gamma=5/3$ if $\rs\to0$.
For a given distribution of entropy $S(\Psi)$, the only degree of freedom
left by Eqs.~(\ref{mpsi}), (\ref{m53}) and (\ref{machaccr}) is the value of
$S_1$, \ie the amount of mass accreted subsonically. The value of $S_{1{\rm
opt}}$ which maximizes the total mass accretion rate satisfies an integral
equation depending only on $S(\Psi)$, which can be solved numerically. 
Thus if the shape of the shock is known, the distribution $S(\Psi)$
is fixed by the Rankine-Hugoniot jump conditions along the shock (see
Appendix~\ref{Aaxis}), and the total mass accretion rate ${\dot M}\{S(\Psi),
S_1\}$ is bounded by this maximum value ${\dot M}\{S(\Psi),S_{1{\rm
opt}}\}$.\\  
Since the subsonic accretion of material with a non--uniform
entropy is possible for $\gamma=5/3$, we only have the inequality 
$\theta_{\rm so}\ge\theta_{\rm c}$. Nevertheless, the difference of entropies
in the  region of subsonic accretion is limited by the Eq.~(\ref{machaccr})
to:  
\begin{equation} 
S_1-S(\varpi=0)\le {5\over2}\log {3\over4}\sim 0.72\;.
\end{equation}
An angular sector of passing flow lines may exist, and is described in
Sect.~\ref{noaccretion}.\\
An estimate of the first order departures from spherical symmetry is obtained
by writing the differential Eq.~(\ref{diffpsi}) to first order in the 
vicinity of $r=0$ (see Appendix~\ref{Adiff}). This provides us with a lower
bound on the departure from sphericity for the function $\beta$, when the
entropy is not uniform:
\begin{equation}
0\le\lim_{r\to0}\left({\p\log\beta\over\p\log r}\right)\le1 \;\;{\rm if}\;\;
S'(\Psi)\ne0\;.\label{dlogbeta}
\end{equation}
As in Sect.~\ref{SSsuper}, this ``bending" of the flow lines enforced by the
entropy gradient could be related to the formation  of long--lived radial
shocks in the supersonic region of flows with a non--uniform entropy, as
observed in numerical simulations (Ruffert~1994b).

\section{Mass accretion rate for a spherically symmetric flow with
non--vanishing kinetic and thermal energies at infinity\label{Sspherical}} 

\subsection{Analytical estimates of the mass accretion rate}

Only two situations are known to have analytical solutions:
\par (i) Spherically symmetric accretion from a gas at rest at infinity
(Bondi~1952). The mass accretion rate of the unique transsonic solution is:
\begin{equation}
{\dot M_{\rm B}}\equiv {4\pi G^2M^2 \rho_\infty\over c_\infty^3}
\left\lbrack{1\over2}
\right\rbrack^{\gamma+1\over2(\gamma-1)}
\left\lbrack{4\over5-3\gamma}
\right\rbrack^{5-3\gamma\over2(\gamma-1)}\;,
\label{MB}
\end{equation}
\par (ii) Axisymmetric accretion from a gas with negligible
temperature (Hoyle and Lyttleton~1939):
\begin{equation}
{\dot M_{\rm HL}}\equiv {4\pi G^2M^2 \rho_\infty\over v_\infty^3}\;.
\label{MHL}
\end{equation}
Other normalizations were also proposed to account for the pressure effects
with more realism, \eg a factor $2$ lower (Bondi \& Hoyle~1944).\\
In order to bridge the analytical gap between the case of spherical accretion
described by ${\dot M}_{\rm B}$ (Eq.~\ref{MB}) and the case of supersonic
accretion from a gas of negligible temperature described by ${\dot M}_{\rm
HL}$ (Eq.~\ref{MHL}), Bondi (1952) proposed a simple interpolation
formula, meant to give the correct order of magnitude. Here we use the
normalization adopted later by Shima \etal (1985) and 
Ruffert \& Arnett~(1994) on the
basis of numerical simulations, which is a factor 2 larger:
\begin{equation} 
{\dot M}_{\rm BH}\equiv{4\pi G^2M^2\rho_\infty \over
(c_\infty^2+v_\infty^2)^{3\over2}} \;.\label{BH}
\end{equation}
It has the advantage of being very simple and giving a roughly
correct limiting behaviour for $v_\infty\to 0$ or $v_\infty\to \infty$. 
However, because it is independent of $\gamma$, it is bound to fail to
reproduce quantitatively the analytical spherical accretion rate ${\dot
M}_{\rm B}$, which varies by a factor $\sim 3$ between $\gamma=5/3$ and
$\gamma=4/3$. Moreover, numerical simulations have shown typical
discrepancies of a factor 2--3 with this formula for various Mach numbers 
(see Fig.~\ref{g5/3} and Fig.~\ref{g4/3}, and also Hunt~1979). 

\subsection{The mass accretion rate of a shocked spherical flow}

An intermediate situation, also analytically tractable, is the
case of a spherical accretion with a constant kinetic energy at infinity:
the gas is ``thrown" towards the accretor, from a reservoir at infinity where
the flow speed is $v_\infty$, the density $\rho_\infty$ and the sound speed
$c_\infty$ are uniform and non--vanishing, \ie the kinetic energy and the
thermal energies are constant at infinity. 
\begin{figure} 
\picture 86.7mm by 37.0mm (pipes) 
\caption[]{The gas is ``thrown" towards the accretor in a spherically
symmetric way. ``Pipes" of constant cross section are used to allow for both 
a constant thermal and kinetic energy at infinity. The accretion radius
$r_{\rm A}$ (full circle) can be smaller than the sonic radius $r_0$
(dotted circle). A spherical shock is represented by the thick dotted
circle.}        \label{pipes}    
\end{figure}  

To this end, we imagine converging radial ``pipes" of constant cross section
(Fig.~\ref{pipes}). They remain separate up to a minimum radius $r_{\rm p}$
at which they merge. For a given distance $r_{\rm p}$, if a stationary
solution exists, the mass accretion rate imposed at infinity would be: 
\begin{equation}
{\dot M}=4\pi  r_{\rm p}^2 \rho_\infty v_\infty \;.
\label{finpipe}
\end{equation}
Thus $2r_{\rm p}$ is also the radius of a cylinder containing the accreted
material in an axisymmetric flow where the transition between the parallel
flow at infinity and the spherical flow near the accretor (and also the
decrease of angular momentum along each flow line) would be artificially
driven through such pipes. This situation allows us to understand how the
kinetic energy influences the mass accretion rate, for a gas with an
adiabatic index $\gamma$.\\ 
For spherical flows, the mass accretion rate of the
stationary solution is determined by the regularity condition at the sonic
radius. We obtain exactly the same dimensionless quartic equation as in
the classic spherical Bondi accretion, determining, in the notations of
Theuns \& David (1992), $\tilde r$ as a function of $\tilde c^2$ (see for
example their Eq.~2.9), except that the velocities $v,c$ and distance $r$ 
are now normalized differently:    
\begin{eqnarray}  
 v&=& \tilde v c_\infty\left(
1+{\gamma-1\over2}\M_\infty^2\right)^{1/2},\\  
r&=& \tilde r {GM\over
c_\infty^2+{\gamma-1\over2}v_\infty^2}\;.
\end{eqnarray}
For spherically symmetric flows, the regularity condition at the sonic
radius fixes the mass accretion rate $\dot M_{\rm sph}$ for the unique
transsonic solution: by integrating the mass flux (\ref{momentum}) over the
sonic sphere ($\M(r_0)=1$), we obtain:  
\begin{eqnarray}
{\dot M_{\rm sph}}&=&4\pi G^2M^2 
\left\lbrack{1\over2}\right\rbrack^{\gamma+1\over2(\gamma-1)}
\left\lbrack{4(\gamma-1)\over5-3\gamma}
\right\rbrack^{5-3\gamma\over2(\gamma-1)}\nonumber\\
& & {\rho_\infty\over c_\infty^{2\over\gamma-1}}
\e^{-\Delta S} B_\infty^{5-3\gamma\over2(\gamma-1)}\;,
\label{mdotsph}
\end{eqnarray}
where $\Delta S$ is the entropy gained when passing through a spherical 
shock, if the flow at infinity is supersonic.\\
Introducing the entropy $S_\infty$ of the gas at infinity defined by
Eq.~(\ref{defentro}), the spherical mass accretion rate can also be written:
\begin{eqnarray} 
{\dot M_{\rm sph}}&=&4\pi G^2M^2 
\left\lbrack{1\over2}\right\rbrack^{\gamma+1\over2(\gamma-1)}
\left\lbrack{4(\gamma-1)\over5-3\gamma}
\right\rbrack^{5-3\gamma\over2(\gamma-1)}\nonumber\\ & & {1\over
\gamma^{1\over\gamma-1}} \e^{-(S_\infty+\Delta S)}
B_\infty^{5-3\gamma\over2(\gamma-1)}\;. 
\label{mdotsph2}
\end{eqnarray}
{\it The mass accretion rate is clearly an increasing function of energy}
$B_\infty$, {\it and a decreasing function of the total entropy}
$(S_\infty+\Delta S)$.\\
We now consider the two possible cases of flows with and without a
shock separately. 

\subsubsection{Flows without a shock}
According to Eq.~(\ref{mdotsph}) when $\Delta S=0$, we expect the mass
accretion rate of isentropic flows to increase with the Mach number:    
\begin{eqnarray}
{\dot M_{\rm sph}}(\Delta S=0)&=&{\dot M_{\rm B}}
\left\lbrack1+{\gamma-1\over2}\M^2_\infty
\right\rbrack^{5-3\gamma\over2(\gamma-1)}\;.
\label{mdotsub}
\end{eqnarray}
This increase of the mass accretion rate with the Mach number for
spherical subsonic flows, is not obvious {\it a priori}. An increase
of the speed at infinity acts as an additional external pressure on the
flow: according to Eq.~(\ref{rspherique}), the sonic radius $r_0$ decreases, 
and Eq.~(\ref{momentum}) shows that the momentum increases. However, the
decrease of the area of the sonic surface, over which the momentum must be
integrated to obtain the mass accretion rate, is more than
compensated by the increase of momentum, for $\gamma<5/3$. The mass 
accretion rate is independent of the Mach number for $\gamma=5/3$.

\subsubsection{Flows with a shock}
Increasing the velocity of the flow at infinity increases its total energy
$B_\infty$, but passing through a shock also increases its entropy.
According to Eq.~(\ref{mdotsph2}), these two effects act in opposite
directions on the mass accretion rate. Whether the mass accretion is
eventually  increased or decreased depends on the position of the shock, 
which can be determined numerically as follows: for high Mach numbers, we 
choose the distance of the pipes, so that the sound velocity prior to the
shock  $c_1$ equals the sound velocity at infinity $c_1=c_\infty$. This
corresponds to the position $r_{\rm p}>r_{\rm sh}$ such that the dilution due
to gravitational acceleration is balanced by the concentration of flow lines.
We make this particular choice by analogy with the case of axisymmetric
accretion with a detached bow shock: we show in  Appendix~\ref{Ahyperbolae}
that the density enhancement, along the axis of symmetry, before the shock,
is smaller than a factor 2 if $r_{\rm sh}>0.03 r_{\rm A}$.\\
For each radius $r$, the Bernoulli equation~(\ref{Bernoulli}) determines the
supersonic speed $v_1(r)$ and Mach number $\M_1(r)$ prior to the shock. Using
the Rankine--Hugoniot jump conditions, we can compute the entropy jump 
$\Delta S(r)$ and the corresponding subsonic velocity $v_2(r)$ after the
shock, if the shock occurs at that radius $r$:      
\begin{eqnarray} 
v_1(r)&=&\left(v_\infty^2+{2GM\over r}\right)^{1/2}= \M_1(r) c_\infty\;,
\label{bernouspher}\\
v_2(r)&=& {2+(\gamma-1)\M_1^2\over (\gamma+1)\M_1^2} v_1(r)\;,\\
\e^{-\Delta S(r)}&=&
\left\lbrack{(\gamma+1)\M_1^2
\over2+(\gamma-1)\M_1^2}\right\rbrack^{\gamma\over\gamma-1}
\left\lbrack{\gamma+1\over2\gamma
\M_1^2-\gamma+1}\right\rbrack^{1\over\gamma-1}\;. 
\label{entroRH}
\end{eqnarray}
For a given entropy $\Delta S$ and Bernoulli
constant $B_\infty$, we denote the velocity profile of the unique transsonic
solution, by $v[\Delta S,B_\infty](r)$. The unique shock position $r_{\rm
sh}$ corresponds to the solution of the equation  
\begin{equation}
v_2(r_{\rm sh})=v[\Delta S(r_{\rm sh}),B_\infty](r_{\rm sh})\;.
\end{equation} 
\begin{figure} 
\picture 86.7mm by 72.3mm (shock) 
\caption[]{Position of the shock $r_{\rm sh}$, in accretion radii, as a
function of $\M_\infty$. The value of $\gamma$ is indicated on each
curve.}         \label{solnum}     
\end{figure}  
The position of the shock determined by this procedure is shown in
Fig.~\ref{solnum}, as a function of $\gamma$ and $\M_\infty$. Except for
values of $\gamma$ close to $5/3$, the spherical shock distance is 
larger than the accretion radius, and thus larger than was observed in
numerical simulations of the axisymmetric BHL flow (typically $r_{\rm
sh}\sim0.2 r_{\rm A}$).\\ The spherical mass accretion rate is given by the
Eq.~(\ref{mdotsub}) diminished  by a factor $\exp(-\Delta S(r_{\rm sh}))<1$
defined in Eq.~(\ref{entroRH}).\\
\begin{figure} 
\picture 86.7mm by 142.0mm (bosse) 
\caption[]{The mass accretion rate, as a function of $\M_\infty$, in
Bondi units ${\dot M}_{\rm B}$ (upper curves) and also in units of the
interpolation formula ${\dot M}_{\rm BH}$ (lower curves). The value of
$\gamma$ is indicated on each curve. The mass accretion rate follows the
Hoyle--Lyttleton behaviour when the kinetic energy dominates the thermal
energy (slope $\M^{-3}$). For a nearly isothermal flow, however, it
exceeds the Bondi value if the thermal energy is larger than the kinetic
energy. The dotted horizontal line is the mass accretion rate from a gas at
rest at infinity.}         \label{bosse}        
\end{figure}   
Increasing the kinetic
energy of the supersonic flow always decreases ultimately the mass accretion
rate below the value ${\dot M}_{\rm B}$ obtained for a gas at rest. Between
the two opposite tendencies (entropy and energy), the decrease due to entropy
will dominate for large Mach numbers, since the effective Mach number $\M_1$
at the shock radius is larger than the Mach number at infinity. For
$\M_\infty\gg1$, Eqs.~(\ref{mdotsph2}) and (\ref{entroRH}) imply:
\begin{equation}  
{\dot M}_{\rm sph} \sim 
C(\gamma)
\left\lbrack{\M_\infty\over \M_1}\right\rbrack^{2\over\gamma-1}
{\rho_\infty\over v_\infty^3}\le
C(\gamma) {\rho_\infty\over v_\infty^3} \;.
\label{asymptot}
\end{equation}
Moreover, Fig.~\ref{solnum} and Fig.~\ref{bosse} show that the shock radius
is proportional to the accretion radius for high Mach numbers, in other
words the ratio $\M_\infty/\M_1$ is a non--vanishing constant for high
Mach numbers. Apart from a multiplicative factor $C(\gamma)$ depending on
$\gamma$, the mass accretion rate in this spherical geometry scales like the
mass accretion rate ${\dot M}_{\rm HL}$ derived by Hoyle \& Lyttleton (1939)
for an axisymmetric flow of negligible pressure. It is remarkable that the
mass accretion rate is asymptotically independent of the value of the
sound speed at infinity.

\subsection{What is the maximum spherical mass accretion rate~?}

We see from Eq.~(\ref{mdotsph2}) that ${\dot M}_{\rm B}$ is the maximum
spherical mass accretion rate for $\gamma=5/3$. As we shall proceed to show,
the situation is very different for $\gamma$ close to 1. The value of 
$\gamma$ defines the strength of a pressure change in response to a density
change. Since the pressure gradient is the only force limiting the mass
accretion rate, the higher the value of $\gamma$,
the higher the  pressure forces and thus the slower the accretion (see
Eq.~\ref{MB}).\\ 
An isothermal gas is much more easily ``swallowed" by the
accretor: a significant increase of the mass accretion rate occurs for
$1<\M_\infty^2<2/(\gamma-1)$. In this range, the thermal energy dominates
the kinetic energy, and Eq.~(\ref{rspherique}) implies that the sonic radius
scales like the Bondi radius $\sim GM/c_\infty^2$. More accurately, taking 
the Rankine--Hugoniot jump condition (\ref{entroRH}) in the limit  $\gamma\to
1$ and combining with Eq.~(\ref{mdotsph2}) gives:   
\begin{equation} 
{\dot M}_{\rm sph}(\gamma\to1)={\dot M}_{\rm B}
\M_1^2\exp\left({\M_\infty^2-\M_1^2\over2}+{1\over2\M_1^2}\right)\;.
\end{equation}
We can estimate the contribution of the exponential factor using the 
Bernoulli equation (\ref{bernouspher}), for $\gamma\to 1$:  
\begin{equation}
0<{\M_1^2-\M_\infty^2\over2}=2{r_0\over r_{\rm sh}}<2\;.
\label{m1minf}
\end{equation} 
Consequently, the mass accretion rate scales like: 
\begin{equation}
{\dot M}_{\rm sph}(\gamma\to1)= {\cal O}( {\dot M}_{\rm B}\M_\infty^2) \;\;
{\rm for}\;\; 1\ll\M_\infty^2<{2\over\gamma-1}\;. 
\label{entroenerg}
\end{equation}
Using the geometrical constraint that $r_{\rm sh}>r_0$, we can also write the
mass accretion rate ${\dot M}_{\rm sh}$ as the flux of mass through the shock
surface, for $\gamma\to 1$. For high Mach numbers, Eq.~(\ref{bernouspher})
ensures that $v_1\sim v_\infty$, and the density
before the shock $\rho_1\sim\rho_\infty$ because the pipes are of constant
cross section. Thus,
\begin{equation}  
{\dot M}_{\rm sh}\equiv 4\pi r_{\rm
sh}^2\rho_1 v_1 =  {4\pi G^2M^2\rho_\infty\over
c_\infty^3}   \left({r_{\rm sh}\over r_0}\right)^2 {\M_\infty\over 4}\;. 
\end{equation}
Comparing this equation to Eq.~(\ref{entroenerg}), we conclude from 
${\dot M}_{\rm sh}={\dot M}_{\rm sph}$ that the
shock radius is a factor $\M_\infty^{1/2}$ larger than the sonic radius.
Inserting this result back into Eq.~(\ref{m1minf}), we further conclude that 
the mass accretion rate is:
\begin{eqnarray}
{\dot M}_{\rm sph}(\gamma\to 1)= \e^{3/2}\pi{\rho_\infty G^2M^2\over
c_\infty^3} \M_\infty^2\;,\nonumber\\
{\rm for}\;\; 1\ll\M_\infty^2<{2\over\gamma-1}\;, 
\end{eqnarray}
which is a factor $\M_\infty^2$ larger than the spherical mass accretion
rate  ${\dot M}_{\rm B}$ from a gas at rest at infinity, and a factor
$\M_\infty^5$ higher than the Hoyle--Lyttleton accretion rate 
${\dot M}_{\rm HL}$.\\ 
We conclude that for values of $\gamma$ approaching 1, the largest
spherical mass accretion rate is:
\begin{equation}
{\dot M}_{\rm max}\sim {2\over\gamma-1}{\dot M}_{\rm B}\;,
\end{equation}
and is reached when the kinetic and thermal energies at infinity are
comparable, \ie: 
\begin{equation}
\M_\infty^2\sim {2\over\gamma-1}\;.
\end{equation}
This scaling obtained in the isothermal limit for high Mach numbers is
confirmed by the exact calculation of the mass accretion rate, for any value
of $\gamma$, through the numerical derivation of the  position of the shock,
as shown by  Fig.~\ref{bosse}.\\  
The entropy--energy approach for a spherical flow has shown us that the mass
accretion rate ultimately decreases like the Hoyle--Lyttleton rate when the
kinetic energy dominates the thermal energy.\\ Nevertheless, the mass
accretion rate increases considerably above its value at $\M_\infty=0$,
when the kinetic energy is comparable to the thermal energy,
\ie $1<\M_\infty^2\sim 2/(\gamma-1)$. This effect is most pronounced for
$\gamma\to1$. Thus the value of the adiabatic index plays a very important
role in the determination of the mass accretion rate in the case of spherical
accretion with a shock. Similarly, we expect the mass accretion
rate of an axisymmetric BHL flow with constant parallel velocities at 
infinity to depend on the adiabatic index.

\section{An interpolation formula for the mass accretion rate of an
axisymmetric flow with $\gamma>9/7$\label{Sinterp}}
 
The axisymmetric mass accretion rate ${\dot M}_{\rm ax}$ can be
formally calculated from the integral of the mass flux along any surface
surrounding the accretor. We can decompose the  mass accretion into three
components,
\begin{equation}
{\dot M}_{\rm ax}={\dot M}_{\rm so}+{\dot M}_{\rm S}+{\dot M}_{\rm F}\;,
\end{equation}
where ${\dot M}_{\rm so}$ is the mass flux through the sonic surface,
${\dot M}_{\rm S}$ is the mass flux in the subsonic region (Types SF and
FSF), and ${\dot M}_{\rm F}$ is the mass flux in the supersonic region ahead
of the shock (Type FSF only).\\
The trajectories of the gas in the supersonic region ahead of the
accretor are well approximated by hyperbolae, obtained by neglecting the
temperature of the gas. ${\dot M}_{\rm F}$ decreases to zero when
the accretor size $\rs$ decreases to zero (Eddington~1926):
\begin{equation}
{\dot M}_{\rm F}\sim\pi \rs(r_{\rm A}+\rs)\rho_\infty v_\infty\;.
\end{equation}
According to Sect.~\ref{Ssubsonic}, ${\dot M}_{\rm S}$ also
decreases to zero with decreasing size of the accretor, for shocked flows 
with $\gamma<5/3$. Consequently, the mass flux through the sonic surface
gives a good estimate of the total mass accretion rate for small accretors:
\begin{equation}
{\dot M}_{\rm ax}={\dot M}_{\rm so}(1+\epsilon(\rs))\;\;{\rm for
}\;\;1<\gamma<5/3\;.
\end{equation}
Using Eq.~(\ref{momentum}) with $\M=1$, we can write
${\dot M}_{\rm so}$ as follows:    
\begin{eqnarray} 
{\dot M}_{\rm so}&\equiv&4\pi G^2M^2 {\rho_\infty\over
c_\infty^{2\over\gamma-1}} \left\lbrack{2(\gamma-1)\over\gamma+1}
\right\rbrack^{\gamma+1\over2(\gamma-1)}\nonumber\\
& &\int_{\rm sonic}
\left\lbrack{1+\bvr_{\rm so}\over
\bvr_{\rm so}^{5-3\gamma\over\gamma+1}}
\right\rbrack^{\gamma+1\over2(\gamma-1)}
B_\infty^{5-3\gamma\over2(\gamma-1)}\e^{-\Delta S}
{\d\Phi\over 4\pi r_{\rm so}^2}\;. 
\label{mdotfr}
\end{eqnarray}
As in the case of spherical accretion, the mass accretion rate is an
increasing function of the energy and a decreasing function of the entropy.
However, the geometry of the sonic surface also influences the mass
accretion rate. We know from Eq.~(\ref{sonic}) that the shape of the sonic
surface $r_{\rm so}(r,\theta)$ is directly related to the shape of the flow
lines. In contrast to the energy and the entropy, the geometrical
factor depends on  the shape of the flow lines at infinity, and is more
difficult to estimate since it does not correspond to a conserved quantity.
One can try to obtain, for the axisymmetric case, an
interpolation formula for the mass accretion rate ${\dot M}_{\rm I}$, based
on the spherical entropy--energy approach.\\ 
With some trigonometry we write:   
\begin{equation}
{\d\Phi\over 4\pi r_{\rm so}^2}=\left(\cos\beta-\sin\beta{\p\log r_{\rm
so}\over \p\theta}\right) {\sin\theta\over 2}\d\theta\;.
\end{equation}
The direction $\theta=0$ being in the direction of the flow, the
axisymmetric sonic surface is described by $r_{\rm so}(\theta)$, with
$0<\theta<\theta_{\rm so}$. For Type F flows, $\theta_{\rm so}$ must be
replaced by $\pi$ in what follows:  
\begin{eqnarray}
{\dot M}_{\rm so}&=&4\pi{\rho_\infty G^2M^2\over c_\infty^3} 
\left\lbrack{1\over2}\right\rbrack^{\gamma+1\over2(\gamma-1)}
\left\lbrack{1+{\gamma-1\over2}\M_\infty^2\over{5-3\gamma\over4}}
\right\rbrack^{5-3\gamma\over2(\gamma-1)} \nonumber \\
&\int_0^{\theta_{\rm so}}&
h(\bvr_{\rm so})
\e^{-\Delta S(\M_1(\theta))}
{\d\Phi\over 4\pi r_{\rm so}^2}\;,\label{mdotfr22}\\
h(\bvr)&\equiv&
\left\lbrack {\bvr_0\over  \bvr}
\right\rbrack^{5-3\gamma\over2(\gamma-1)}
\left\lbrack {1+\bvr\over 1+\bvr_0 }
\right\rbrack^{\gamma+1\over2(\gamma-1)}\;.
\end{eqnarray}
First note by differentiating $h(\bvr)$ with respect to $\bvr$ that:
\begin{equation}
h(\bvr)\ge h(\bvr_0)=1\;,\label{mini}
\end{equation}
The integral in Eq.~(\ref{mdotfr22}) can be seen as an average of
$h(\bvr_{\rm so})$ times the entropy weighted by the geometrical function
$\d\Phi/ 4\pi r_{\rm so}^2>0$, thus defining an effective entropy
corresponding to an effective Mach number $\M_{\rm e}\equiv \M_1(\theta_{\rm
e})$:    
\begin{eqnarray}
&\int_0^{\theta_{\rm so}}&
h(\bvr_{\rm so})
\e^{-\Delta S(\M_1(\theta))}
{\d\Phi\over 4\pi r_{\rm so}^2}\nonumber\\
&\equiv&
h(\bvr_{\rm so}(\theta_{\rm e}))
\e^{-\Delta S(\M_e)}
\int_0^{\theta_{\rm so}}{\d\Phi\over 4\pi r_{\rm so}^2}\;.
\label{approxme}
\end{eqnarray} 
Using Eq.~(\ref{entroRH}) and the definition~(\ref{MB}) of ${\dot M}_{\rm
B}$, Eq.~(\ref{mdotfr22}) becomes:      
\begin{eqnarray}
{\dot M}_{\rm so}&=&{\dot M}_{\rm B}\left\lbrack1+{\gamma-1\over2}\M^2_\infty
\right\rbrack^{5-3\gamma\over2(\gamma-1)}h(\bvr_{\rm so}(\theta_{\rm e}))
\int_0^{\theta_{\rm so}}
{\d\Phi\over 4\pi r_{\rm so}^2} \nonumber\\
& &\left\lbrack{(\gamma+1)\M_{\rm e}^2
\over2+(\gamma-1)\M_{\rm e}^2}\right\rbrack^{\gamma\over\gamma-1}
\left\lbrack{\gamma+1\over2\gamma
\M_{\rm e}^2-\gamma+1}\right\rbrack^{1\over\gamma-1}\;.
\label{maxme}
\end{eqnarray}
One can check that for isentropic flows ($\Delta S=0$) and small deviations
from sphericity ($f(\theta)\sim \beta\sim (\bvr_{\rm so}-\bvr_0) \ll 1$), the
mass accretion rate is accurately described by the spherical model of
Sect.~\ref{Sspherical}:   
\begin{equation}
{\dot M}_{\rm so}={\dot M}_{\rm sph}(\M_\infty)
\left\lbrack 1+{\cal O}\left\lbrace {\rm Max}(\bvr_{\rm so}-\bvr_0)^2
\right\rbrace\right\rbrack\;.
\label{mdotfr3}
\end{equation}
In order to approximate the mass accretion rate ${\dot M}_{\rm so}$, one must
estimate the value of the effective Mach number $\M_{\rm e}$, the effective
sonic radius $\bvr_{\rm so}(\theta_e)$ and the integral of 
$(\d\Phi/ 4\pi r_{\rm so}^2)$. 
If the shock is detached, we can make simple guesses if the
spherical sonic radius $r_0$ is smaller than the accretion radius $r_{\rm
A}$. We then use the spherical approximation for $\Phi$ and $r_{\rm so}$, and
use an ad-hoc expression for $\M_{\rm e}$ only, to account for the asymmetry.
We see from Fig.~\ref{r0rA} that $r_0<r_{\rm A}$ for $5/3>\gamma>9/7$. The
range $9/7>\gamma>1$, and especially the vicinity of $\gamma=1$, requires a
more careful treatment because $r_0$ can be orders of magnitude larger than
$r_{\rm A}$. Consequently, we restrict the present approach to the range
$5/3>\gamma>9/7$, and make the simple approximation $\bvr_{\rm
so}(\theta_e)\sim \bvr_0$.\\ 
Since we would like to recover the
Hoyle--Lyttleton accretion rate  ${\dot M}_{\rm HL}$ for high values of the
Mach number, we see from Eq.~(\ref{asymptot}) that this requires a constant
value for the ratio $\M_{\rm e}/\M_\infty$ when $\M_\infty\to\infty$. As a
first approximation, we propose the ad--hoc prescription that ${\dot M}_{\rm
so}\to\lambda {\dot M}_{\rm HL}$ when $\M_\infty\to\infty$, and retain the
same ratio $\M_{\rm e}/\M_\infty$ for intermediate values of $\M_\infty$.
Neglecting the variations of the sound velocity before the shock, we can
write the effective Mach number in terms of the angle $\varphi_{\rm e}$ of
the ``effective flow line" of entropy $S(\M_{\rm e})$ and the normal to
the shock surface, and the distance $r_{\rm e}$ between the accretor and the
intersection of this flow line with the  shock surface:
\begin{equation}
{\M_{\rm e}\over\M_\infty}=\cos\varphi_{\rm e}\left(1+{r_{\rm A}\over 
r_{\rm e}} \right)^{1\over2}\;.\label{varphie}
\end{equation}
A geometrical interpretation is the following. Let the accretion
radius $r_{\rm A}$ act as a scaling factor on the topology of the accreted
flow, and keep the angle $\varphi_{\rm e}$ and the
ratio $r_{\rm A}/r_{\rm e}$ independent of the Mach number.
Eq.~(\ref{varphie}) then implies that the ratio $\M_{\rm
e}/\M_\infty$ is constant. This does not contradict
the fact that the opening angle of the Mach cone, well after the accretor,
depends on the Mach number, since our approximation deals with the region of
the bow shock which is mainly ahead of the accretor.\\  
We obtain, for $\M_{\rm e}\ge 1$:        
\begin{eqnarray}
&&\int_0^{\theta_{\rm so}}{\d\Phi\over 4\pi r_{\rm so}^2}\sim 1\;,\\
&&\bvr_{\rm so}(\theta_e)\sim \bvr_0\;,\\ {\M_{\rm
e}\over\M_\infty}&\sim&{1\over2^\gamma\lambda^{\gamma-1\over2}}
\left\lbrack{2\over\gamma}\right\rbrack^{1\over2}
{(\gamma+1)^{\gamma+1\over2}\over(\gamma-1)^{5(\gamma-1)\over4}
(5-3\gamma)^{5-3\gamma\over4}}\;.\label{meminf}  
\end{eqnarray}
With $\M_{\rm e}$ defined by this relation if $\M_{\rm e}\ge 1$, and 
$\M_{\rm e}= 1$ otherwise, our interpolation formula follows from
Eq.~(\ref{maxme}):      
\begin{eqnarray}
{\dot M}_{\rm I}&\equiv&{\dot M}_{\rm
B}\left\lbrack1+{\gamma-1\over2}\M^2_\infty
\right\rbrack^{5-3\gamma\over2(\gamma-1)} \nonumber\\
& &\left\lbrack{(\gamma+1)\M_{\rm e}^2
\over2+(\gamma-1)\M_{\rm e}^2}\right\rbrack^{\gamma\over\gamma-1}
\left\lbrack{\gamma+1\over2\gamma
\M_{\rm e}^2-\gamma+1}\right\rbrack^{1\over\gamma-1}\;.
\label{interpol}
\end{eqnarray}

\begin{figure} 
\picture 86.7mm by 72.3mm (i5/3) 
\caption[]{Comparison between our interpolation formula and the
numerical simulations. Mass accretion rates are in ${\dot M}_{\rm BH}$ units.
The solid line corresponds to $\lambda=1$, the dotted line to $\lambda=0.5$.
For $\gamma=5/3$, the numerical simulations provide only upper bounds to the
mass accretion rate of a point mass because the sonic surface is always
attached to the accretor.}          \label{g5/3}       
\end{figure}  

\begin{figure} 
\picture 86.7mm by 72.3mm (i4/3) 
\caption[]{Same as Fig.~\ref{g5/3}, but for $\gamma=4/3$.}        
\label{g4/3}    
\end{figure}  
We extend our rough approximation to the case $\gamma=5/3$, neglecting
the mass accreted in the subsonic region. We know from Sect.~\ref{S53} that
this corresponds to a small (but not zero) fraction of the total flux of
mass.\\ 
The curves displayed in Fig.~\ref{g5/3} and Fig.~\ref{g4/3} show the mass 
accretion rates obtained with our interpolation formula
${\dot M}_{\rm I}$, the Bondi--Hoyle interpolation formula ${\dot M}_{\rm
BH}$ and the results of numerical simulations 
(Ruffert~1994b, 1995; Ruffert \& Arnett~1994). 
For $\gamma=5/3$ our interpolation Eq.~(\ref{interpol}), in ${\dot M}_{\rm
BH}$ units, rises monotonously with Mach number (Fig.~\ref{g5/3}). The
numerical results fit only marginally well. Note that the numerical
simulations with $\gamma=5/3$ give only an upper bound for the mass accretion
rate, since the accretor is always larger than the spherical sonic radius. 
For $\gamma=4/3$ (Fig.~\ref{g4/3}) a prominent maximum is apparent both on 
the interpolation formula and on the numerical results. 
Comparing Fig.~\ref{g4/3} with the corresponding curve in
the lower part of Fig.~\ref{bosse} gives an indication on the role played  
by non--sphericity in reducing the mass accretion rate.\\
Despite the considerable simplification of ignoring the
geometrical term $\Phi(\theta)$ and keeping the ratio $\M_{\rm e}/\M_\infty$
independent of $\M_\infty$, these curves are in reasonable agreement
with the numerical simulations for $\gamma=4/3$ and $\gamma=5/3$. Thus we
expect this analytic formula to be relevant for the whole range
$5/3<\gamma<9/7$.\\ 
Note that we do not solve the question of the dependence
of the mass accretion rate with respect to the adiabatic index for high Mach
numbers. For this purpose, and following Bondi \& Hoyle (1944), we introduced
the constant $\lambda$, assumed to be between $0.5$ and $1$. Our
interpolation formula is nothing more than a natural way to link the Bondi
mass accretion rate at $\M_\infty=0$ with the Hoyle-Lyttleton mass accretion
rate at $\M_\infty\to\infty$. \\ One should also keep in mind that the mass
accretion rate obtained in the numerical simulations is for most models a
time average over the fluctuations or quasi periodic oscillations due to the
instability of the flow, and therefore differs from the exact mass accretion
rate of the solution of the stationary equations. This instability is
stronger for $\gamma$ close to $5/3$. 

\section{Conclusions \label{Sconclusion}}

By using a geometrical interpretation of the shape of the flux tube at the
sonic radius, we have shown the particular significance of the radius
$r_0$ (Eq.~\ref{rspherique}), which depends only on the adiabatic index and
the energy of the flow. It is not only the sonic radius of all spherical
flows, but is also related through the property ${\cal P}$
(Sect.~\ref{SSprop}) to the sonic surface of axisymmetric flows.\\  
By reducing the stationary flow equations to a single partial
differential equation, we have performed a local analysis near the accretor
and have obtained some general results which aim at clarifying the diversity
of possible configurations. We have extended the classification introduced by
Bondi (1952) to axisymmetric stationary flows. In addition to the
subsonic and supersonic types of accretion, angular sectors without any
accretion are also possible, as well as singular directions of infinitely
high accretion flux. This is illustrated by the self-similar solution found
by Bisnovatyi-Kogan \etal (1979), which is a very particular singular case
($\theta_{\rm c}=0$).\\  
The angular sectors without any accretion is subsonic for flows with a
detached shock, so that $\theta_{\rm so}\le\theta_{\rm c}$ in the notation
introduced in Sect.~\ref{notations}.\\ 
Among some of the properties that we established for flows with
$\gamma<5/3$ are the following: 
\par(i) the sonic surface is likely to be
attached to the accretor for nearly isothermal flows with
high Mach numbers, 
\par(ii) isentropic accretion proceeds mostly from the downstream
hemisphere of the accretor,
\par(iii) subsonic accretion must be isentropic.\\
We have also found a series of properties for flows with $\gamma=5/3$:
\par(i) accretion is always regular, 
\par(ii) the sonic surface is always attached to the accretor, 
\par(iii) the pressure distribution is spherically symmetric to first order, 
\par(iv) the Mach number and the entropy at a point like
accretor are algebraically related to the local mass flux,
\par(v) shocked matter is accreted mostly from the
downstream hemisphere, where the density, velocity and Mach number are
maximum, and the temperature and entropy are minimum,
\par(vi) isentropic accretion is spherically symmetric to first order for the
velocity, temperature, density, mass flux, and Mach number.\\
 
3--D numerical simulations have shown that the strongest instabilities
occur with the adiabatic index $\gamma=5/3$. The fact that subsonic regions
invariably reach the accretor for $\gamma=5/3$ might be considered as a
warning about the possible role, in the instability mechanism, of the 
boundary conditions at the surface of the accretor. For $\gamma<5/3$ and
sufficiently small accretors, it would be interesting to check for the
presence of the instability in cases where the accretor is surrounded by a
supersonic region (Type~F in the nomenclature of Sect.~\ref{SStypes}).\\

We have also determined the departure from sphericity, and constrained the 
radial variation of the angle $\beta$ for $r\to0$. In particular, we have
stressed the influence of the entropy gradient on the bending of the flow
lines for supersonic flows with $\gamma<5/3$ and $\gamma=5/3$.\\

We have stressed the role of the geometrical shape of the shock in the
determination of the entropy of the accreted matter, and contrasted it with
the simpler case of shocked spherical flows. In this latter case, we have
shown the relevance of the entropy-energy approach, and noted that the mass
accretion rate scales like the Hoyle--Lyttleton formula for high Mach
numbers. Moreover, we have outlined the particularity
of nearly isothermal flows for which the mass accretion rate can be a factor
$2/(\gamma-1)$ higher than the Bondi mass accretion rate, when the kinetic
energy at infinity is comparable to the thermal energy. This demonstrates the
fundamental difference of the BHL flows that have $\gamma$ close to 1 and
those with $\gamma$ close to 5/3. \\  
We have shown that the total mass accretion rate equals the mass accretion
rate through the sonic surface, for shocked regular flows with $\gamma<5/3$.
We have derived an interpolation formula for the mass accretion rate valid in
the range $9/7<\gamma<5/3$. It links spherical accretion to the case of
highly supersonic accretion, with reasonable agreement with numerical
simulations.\\

Analytical estimates of the following quantities, however, are still missing:
\par(i) the distance of the shock as a function of $\gamma$ and $\M_\infty$,
\par(ii) the value of $\theta_{\rm so}$ and $\theta_{\rm c}$ as a function of
$\gamma$ and $\M_\infty$,
\par(iii) the mass accretion rate in the limit of infinite Mach numbers, as a
function of $\gamma$.

Nevertheless, our classification will help perform, in a
forthcoming paper, a local stability analysis of all the configurations
considered.

\acknowledgements
TF was supported by the EC grant ERB-CHRX-CT93-0329, as part of the research
network `accretion onto compact objects and protostars'. We acknowledge
useful comments by Dr. H. Spruit and Dr. U. Anzer.

\appendix

\section{Third order partial differential system for an axisymmetric
stationary flow \label{Apartial3}}

The combined vorticity equation, continuity equation and entropy
equation are written as functions of $\M,\beta,S$ as follows:
\begin{eqnarray}
&&(1-\M^2\cos^2\beta){\p\beta\over\p\log r}-
\M^2\sin\beta\cos\beta{\p\beta\over\p\theta}=\nonumber\\
&&
{1-\M^2\over2+(\gamma-1)\M^2}{\p\log\M^2\over\p\theta}
+{1-\M^2\over\gamma\M^2}{\p S\over\p\theta}\nonumber\\
&&+{\sin^2\beta\over\tan\theta}-{\sin\beta\cos\beta\over1+\bvr}
\left\lbrack{2-\gamma\over\gamma-1}-{1+2\bvr\over2}\M^2-\bvr\right\rbrack\;,
\label{dif1}\\
&&{1-\M^2\cos^2\beta\over2+(\gamma-1)\M^2}{\p\log\M^2\over\p\log r}-
{\M^2\sin\beta\cos\beta\over2+(\gamma-1)\M^2}{\p\log\M^2\over\p\theta}=
\nonumber\\
&&-{\p\beta\over\p\theta}+{\tan\beta\over\gamma\M^2}
{\p S\over\p\theta}+{5-3\gamma-2(2-\gamma)\sin^2\beta\over2(1+\bvr)
(\gamma-1)}\nonumber\\
&&-{\bvr\over1+\bvr}
(2\cos^2\beta+1)-{\sin\beta\cos\beta\over\tan\theta}\;,\label{dif2}\\
&&{\p S\over\p\log r}=-\tan\beta{\p S\over\p\theta}\;.\label{dif3}
\end{eqnarray}

\section{Proof of the property ${\cal P}$ of the axisymmetric
sonic surface of Type F flows \label{Aproof}}

Using the axisymmetry condition and some simple trigonometry, the variation 
of the cross--section $\Phi$ along the flow line can be written:     
\begin{equation}
{\p\log\Phi\over\p l}{\p l\over\p\log r}=2+{\tan\beta\over\tan\theta}
+{r\over\cos\beta}{\p\beta\over\p {\bf n}}\;,
\label{trigo}
\end{equation}
where $\beta(r,\theta)$ is defined by Eq.~(\ref{defbeta}), and its derivative
is taken in the direction ${\bf n}$ perpendicular to the flow line (oriented
towards increasing $\theta$). For $\beta=0$ we recover the spherical case of
conical flow tubes, with a cross section proportional to the square of the
distance.\\   
We can use Eq.~(\ref{trigo}) along the detached sonic surface,
together with Eq.~(\ref{sonic}), for $0\le\theta\le\pi$:   
\begin{eqnarray} 
\bvr_{\rm so}&=&{\bvr_0-f(\theta)\over
1+f(\theta)}\;, {\rm with} \label{sonigeom}\\
f(\theta)&\equiv&{1\over2}\left({r_{\rm so}\over\cos\beta}{\p\beta\over\p 
{\bf n}}+ {\tan\beta\over\tan\theta}\right)\;. \label{deff}
\end{eqnarray}
Let us now prove that the sonic radius reaches the value of the spherical
radius at least once, \ie that $f(\theta)$ has to vanish.\\
Axisymmetry imposes $\beta(r,\theta=0)=\beta(r,\theta=\pi)=0$, and by
Eq.~(\ref{deff}): 
\begin{eqnarray}
f(0)&=&{\p\beta\over\p\theta}(0)\;,\label{f0}\\
f(\pi)&=&{\p\beta\over\p\theta}(\pi)\;.\label{fpi}
\end{eqnarray}
Suppose, for example, that the sonic surface is inside the sphere of radius 
$r_0$ and detached from the accretor, \ie $f(\theta)>\epsilon>0$ for
$\theta\in [0,\pi]$. Equations (\ref{f0}) and (\ref{fpi}) imply that
$\beta(r,\theta)$ is strictly positive near $r_{\rm so}(\theta=0)$, and
strictly negative near $r_{\rm so}(\theta=\pi)$. A line must exist in
the ($r,\theta$) plane separating $r_{\rm so}(0)$ from $r_{\rm so}(\pi)$,
along which $\beta$ vanishes, when decreasing from a region of positive
values to the region of negative values. We denote by $r_{\rm so}(\theta_{\rm
i})$ the intersection of this line with the sonic surface: 
\begin{equation} 
f(\theta_{\rm i})={1\over2}\left({r_{\rm so}\over\cos\beta}{\p\beta\over\p
{\bf n}}\right)\le 0\;,
\end{equation}
which contradicts the hypothesis. The same contradiction arises if we suppose
that $f(\theta)<-\epsilon<0$. Consequently $f$ must vanish for some value of
$\theta$.

\section{Linearized partial differential equation for $\gamma< 5/3$
 \label{Ainf53}}

By using Eq.~(\ref{tanbeta}), we can relate the second radial derivative of
$\Psi$ to its first radial derivative, at $r=0$, through the
following equation: 
\begin{equation}
\lim_{\bvr\to0}\left\lbrack
{\bvr^2{\p^2\Psi\over\p\bvr^2}\over
\bvr{\p\Psi\over\p\bvr}} \right\rbrack= \lim_{\bvr\to0}
\left\lbrack{\p\log\beta\over\p\log\bvr}-1+\bvr{\p^2\Psi\over\p\bvr\p\theta}
\left({\p\Psi\over\p\theta}\right)^{-1}\right\rbrack\;.
\label{secondprem}
\end{equation}
The local mass flux is finite for regular flows, and therefore the limit
when $r\to0$ of ($\bvr\p^2\Psi/\p\bvr\p\theta$) is zero in the angular 
sectors where the flow is regular.\\
Let us define the new radial variable $z\equiv r^{(5-3\gamma)/2}$.
Equation~(\ref{implicit}) implies to first order: 
\begin{equation}
z\M^2\sim\left({1\over\sin\theta}{\p\Psi_0\over\p\theta}\right)^{1-\gamma}
\;.\label{zm2}
\end{equation}
\par(i) If Eq.~(\ref{betamax}) is a strict inequality, Eq.~(\ref{Eulert})
implies that the pressure is spherically symmetric to first order. Using
Eqs.~(\ref{mpsi}) and (\ref{Psup}), the stream function $\Psi_0(\theta)$ at
the surface of the accretor is directly related to the entropy $S_0(\theta)$
by:  
\begin{equation}
{\p\over\p\theta}\left\lbrack\left({1\over\sin\theta}
{\p\Psi_0\over\p\theta}\right)^\gamma\e^{-S_0}\right\rbrack=0\;.
\end{equation}
The leading terms of the differential Eq.~(\ref{diffpsi}) satisfy the
following equation:
\begin{eqnarray}
\left\lbrace
(1+\epsilon_1)a{\p^2\over\p \theta^2}+(1+\epsilon_2)b{\p\over\p \theta}+
(1+\epsilon_3)c\right.+\nonumber\\
\left.(1+\epsilon_4)z{\p^2\over\p
z^2}+{3(2-\gamma)\over5-3\gamma}(1+\epsilon_5){\p\over\p
z}\right\rbrace\Psi_1= (1+\epsilon_6)d z\;,\label{linearinf53} 
\end{eqnarray} 
where the six functions $\epsilon_i(z,\theta)$ converge to zero when
$z\to 0$. The functions $a,b,c,d$ depend on $\theta$ only, and are
defined as follows: 
\begin{eqnarray}
&a&\equiv-\left({2\over5-3\gamma}\right)^2
\left({1\over\sin\theta}{\p\Psi_0\over\p\theta}\right)^{\gamma-1}\;,\\
&b&\equiv\left({2\over5-3\gamma}\right)^2
\left({1\over\sin\theta}{\p\Psi_0\over\p\theta}\right)^{\gamma-1}\nonumber\\
&&\left({1\over\tan\theta}+2{\p\Psi_0\over\p\theta}{S'(\Psi_0)\over\gamma}
\right)\;,\\
&c&\equiv\left({2\over5-3\gamma}\right)^2
\left({1\over\sin\theta}{\p\Psi_0\over\p\theta}\right)^{\gamma-1}
\left({\p\Psi_0\over\p \theta}\right)^2
{S''(\Psi_0)\over\gamma}\;,\\
&d&\equiv\left({2\over5-3\gamma}\right)^2
\left({\p\Psi_0\over\p\theta}\right)^{2\gamma}
{S'(\Psi_0)\over\gamma(\sin\theta)^{2(\gamma-1)}}\;.
\end{eqnarray}
Assuming $\lim_{z\to0}(\p\log\Psi_1/\p\log z)>2$ and $d\ne0$ contradicts
Eq.~(\ref{linearinf53}). Thus we conclude that:
\begin{equation}
\left({\p\log\Psi_1\over\p\log z}\right)\le2\;\;{\rm for}\;S'(\Psi_0)\ne0\;.
\end{equation}
By derivating Eq.~(\ref{diffpsi}) with respect to $r$ and considering the
leading terms, we find that $z\p\Psi_1/\p z$ satisfies the same 
Eq.~(\ref{linearinf53}) as $\Psi_1$ does. Using Eq.~(\ref{tanbeta}), we
obtain Eq.~(\ref{betamin}) for regular sectors with a non--uniform entropy.
\par(ii) If Eq.~(\ref{betamax}) is an equality, we use 
Eq.~(\ref{secondprem}) in order to write the first order terms of
Eq.~(\ref{diffpsi}) as follows:   
\begin{equation} 
\lim_{z\to0}{\p\Psi\over \p z}=
{4\e^{S_0}\sin\theta\over3\gamma(5-3\gamma)(2-\gamma)}
{\p\over\p\theta}\left\lbrack\left({1\over\sin\theta}
{\p\Psi_0\over\p\theta}\right)^\gamma\e^{-S_0}\right\rbrack\;.\label{SbMlies}
\end{equation}
In both cases (i) and (ii), $S_0,\Psi_0$ and the radial derivative
$(\p\Psi/\p z)_0$ cease to be a set of three free functions of $\theta$ on 
the surface of the accretor, when the accretor size decreases to zero. Let us
recall that for spherical flows, one parameter (the mass flux) is required as 
boundary condition on the surface $r=\rs$ of the accretor, with no particular
change if $\rs\to 0$. In this respect, the differential system
(\ref{diffpsi}) is more singular at $r=0$ for axisymmetric flows than for
spherical flows.\\

\section{Linearized partial differential equation for $\gamma=5/3$
 \label{Adiff}}

Because the stream function $\Psi$ is finite everywhere, $\epsilon_\Psi$ and
$\bvr^2\p^2\Psi/\p\bvr^2$ tend to zero with decreasing $\bvr$.
Eq.~(\ref{implicit}) implies that when $\gamma=5/3$, the Mach number tends to
a finite limit everywhere where the local mass flux (\ie $\p\Psi_0/\p\theta$ 
according to Eq.~(\ref{mpsi})) is not zero. Taking the limit of
Eq.~(\ref{diffpsi}) when $\bvr\to0$, we find that its last term must vanish
at $\bvr=0$: 
\begin{equation}
{1\over\tan\theta}{\p\Psi_0\over\p\theta}-{\p^2\Psi_0\over\p\theta^2}
+\left({\p\Psi_0\over\p\theta}\right)^2\left(1-{1\over\M_0^2}\right)
{3S'(\Psi_0)\over5}=0\;,
\end{equation}
where the subscript $0$ denotes the value of the function at $r=0$.
This equation is equivalent to the condition that the pressure is spherically
symmetric to first order.\\
In order to find the behaviour of the quantities $\Psi,\M^2$ near $\bvr=0$,
we consider only the leading terms of Eqs.~(\ref{implicit})
and (\ref{diffpsi}). We first define $\M_1$ and $\Psi_1$ as
follows: 
\begin{eqnarray}
\M^2(r,\theta)&\equiv& \M_0^2(\theta)(1+\M_1(r,\theta))\;,\\
\Psi(r,\theta)&\equiv& \Psi_0(\theta)+\Psi_1(r,\theta)\;.
\end{eqnarray}
By linearizing Eq.~(\ref{implicit}) near $\M_0^2\ne1$, we obtain to first
order: 
\begin{equation}
\M_1\sim{4\over3}{3+\M_0^2\over1-\M_0^2}\left\lbrack
-\bvr+{(3+\M_0^2)^4\over2\M_0^2\sin^2\theta}{\p\Psi_0\over\p\theta}
{\p\Psi_1\over\p\theta}\right\rbrack\;.
\end{equation}
Introducing this into Eq.~(\ref{diffpsi}), we obtain the following leading
terms:
\begin{eqnarray}
\left\lbrace
(1+\epsilon_7)e{\p^2\over\p \theta^2}+(1+\epsilon_8)f{\p\over\p \theta}+
(1+\epsilon_9)g\right.+\nonumber\\
\left.(1+\epsilon_{10})\bvr^2{\p^2\over\p
\bvr^2}+{3\over2}(1+\epsilon_{11})\bvr{\p\over\p \bvr}\right\rbrace\Psi_1=
(1+\epsilon_{12})h\bvr\;,\label{linear53} 
\end{eqnarray} 
where the six functions $\epsilon_j(\bvr,\theta)$ converge to zero when
$\bvr\to0$. The functions $e,f,g,h$ depend on $\theta$ only, and are
defined as follows: 
\begin{eqnarray}
&e&\equiv{1\over1-\M_0^2}\;,\\
&f&\equiv-{1\over1-\M_0^2}{1\over\tan\theta}-\nonumber\\
&6&{\p\Psi_0\over\p\theta}
{S'(\Psi_0)\over5\M_0^2}\left\lbrack
{(3+\M_0^2)^5\over3\M_0^2(1-\M_0^2)^2\sin^2\theta}
\left({\p\Psi_0\over\p \theta}\right)^2-1\right\rbrack\;,\\
&g&\equiv\left({\p\Psi_0\over\p \theta}\right)^2
{3S''(\Psi_0)\over5\M_0^2}\;,\\
&h&\equiv-{4\over5}{3+\M_0^2\over(1-\M_0^2)^2}\left({\p\Psi_0\over\p
\theta}\right)^2 {S'(\Psi_0)\over\M_0^2}\;.
\end{eqnarray}
Let us suppose that $l$ is a positive number such that
\begin{equation}
\lim_{r\to0}{\p\log\Psi_1\over\p\log r}=1+l>1\;.\label{hypolim}
\end{equation}
If the entropy is not uniform ($h\ne0$), the first two
terms of Eq.~(\ref{linear53}) are negligible compared to the rhs term.
Dividing the remaining equation by $\bvr$, we obtain:  
\begin{equation}
\left\lbrace (1+\epsilon_7)e{\p^2\over\p \theta^2}+
(1+\epsilon_8)f{\p\over\p
\theta}+ (1+\epsilon_9)g\right\rbrace{\Psi_1\over\bvr}= 
(1+\epsilon_{12})h\;.\label{psi1r}
\end{equation}
According to Eq.~(\ref{hypolim}), $\lim_{r\to0}\Psi_1/\bvr=0$.
Thus Eq.~(\ref{psi1r}) contradicts our hypothesis (\ref{hypolim}), and we
conclude that if $h\ne0$,
\begin{equation}
0\le\left(\lim_{r\to0}{\p\log\Psi_1\over\p\log
r}\right)\le1\;.\label{conclulim} 
\end{equation}
By derivating Eq.~(\ref{diffpsi}) with respect to $\bvr$, and keeping the
leading terms, $\bvr\p\Psi_1/\p\bvr$ satisfies exactly the same equation
(\ref{linear53}) as $\Psi_1$ does. From the same reasoning, we conclude that 
\begin{equation}
0\le\left(\lim_{r\to0}{\p\log\over\p\log
r}{\p\Psi_1\over\p\log\bvr}\right)\le1\;.\label{conclulim2} 
\end{equation}
This lower bound on the scaling of $\bvr\p\Psi_1/\p\bvr$ is also
a lower bound on the scaling of $\beta$: we obtain Eq.~(\ref{dlogbeta}) by
using Eqs.~(\ref{tanbeta}), (\ref{conclulim2}) and the regularity of the
flow for $\gamma=5/3$.

\section{Approximation of the orbits within the supersonic region by
hyperbolae  \label{Ahyperbolae}}

We check the validity of
neglecting the pressure forces in the supersonic flow ahead of the
shock, and assuming that the sound velocity is uniformly
constant to its value at infinity $c_\infty$. The velocity of the
supersonic incoming gas is then defined by the unique hyperbolic
trajectory passing trough this point, with velocity at infinity $v_\infty$.\\
For each point ($r,\theta$) of the shock surface, we can compute numerically
the dimensionless impact parameter $\ba$ (normalized to the
accretion radius $r_{\rm A}$) of this hyperbola: 
\begin{equation}
\br={2\ba^2\over
1+(1+4\ba^2)^{1/2}\cos(\theta-\theta_\infty)} \;.
\label{hyperbole}
\end{equation}
The half angle $\theta_\infty$ of the hyperbola is defined by
\begin{equation}
\cos\theta_\infty={1\over (1+4\ba^2)^{1/2}}\;.
\label{demiangle}
\end{equation}
Let us calculate the variation of density along the axis of symmetry, due to
both the compression by gravitational convergence of the hyperbolic
trajectories and the dilation due to acceleration. This can be done
using the continuity equation~(\ref{continuity}) in cylindrical
coordinates, along the axis, for $\theta\to\pi$:
\begin{equation}
{\p \rho \br^2 v_r\over \p \br}+2\rho \br{\p v_\theta\over\p\theta}=0\;.
\label{contax}
\end{equation}
The azimuthal derivative of the azimuthal component of the velocity is
directly related to the impact parameter $a$ through the conservation of
angular momentum, while the (negative) radial component of the velocity, 
along the axis, is given by the conservation of energy:
\begin{eqnarray}
\br v_\theta&=&-\ba v_\infty\;,\label{angul}\\
v_r(\theta=\pi)&=&-v_\infty\left(1+{1\over
\br}\right)^{1\over2}\;.\label{enervr} 
\end{eqnarray}
In order to determine the azimuthal dependence of the impact parameter, let
us consider the trajectories defined by Eq.~(\ref{hyperbole}) for
$(\theta-\pi)\to0$. Since $a(\pi)=0$, we write:
\begin{equation}
\ba(\theta)=(\theta-\pi){\p \ba\over\p\theta}(\pi)+{\cal
O}\left((\theta-\pi)^3\right)\;. 
\end{equation}
From first order expansions of Eq.~(\ref{demiangle}) and 
Eq.~(\ref{hyperbole}), we find, for $(\theta-\pi)\to0$:
\begin{equation}
\left({\p \ba\over\p\theta}\right)^2+\br{\p \ba\over\p\theta}-{\br\over4}=0
\;.\label{diffba}
\end{equation}
Choosing the negative solution of this equation ($\ba$ is positive and
decreases to zero when $\theta$ increases to $\pi$), leads to:
\begin{equation}
{\p v_\theta\over\p\theta}(\pi)={v_\infty\over2}\left\lbrack1+
\left(1+{1\over\br}\right)^{1\over2}\right\rbrack\;.\label{dvtheta}
\end{equation}
Using Eqs.~(\ref{dvtheta}) and~(\ref{enervr}) in Eq.~(\ref{contax}) we 
obtain: 
\begin{equation} 
{\p\log\rho\over\p \br}(\pi)={1\over
\br^{1\over2}(\br+1)^{1\over2}} -{1\over2}{1+2\br\over\br(1+\br)}\;.
\end{equation}
This can be integrated to obtain the radial dependence of the density along
the axis of symmetry:
\begin{equation}
{\rho\over\rho_\infty}(\pi)={1\over2}+{2+{1\over \br}\over 4\left(
1+{1\over\br}\right)^{1/2}}\;.\label{gravcomp}
\end{equation}
The compression effect due to the convergence of the flow lines is
consequently stronger than the dilatation due to the acceleration of the
flow.\\ 
If the shock is not too close to the accretor, gravitational focusing 
is negligible and the density and temperature can be approximated as
constant in the upstream supersonic flow. One can check from
Eq.~(\ref{gravcomp}) that this density enhancement is smaller than a
factor of two for $\br>0.03$. Note that when the shock is detached, its
distance is generally of order $\br_{\rm sh}(\pi)\sim0.2$ in numerical
simulations. We deduce from Eq.~(\ref{gravcomp}) that the sound velocity
scales like $\br^{(\gamma-1)/2}$ when $\br\to0$, and is therefore
always negligible compared to the gravitational potential scaling like
$\br^{-1}$.

\section{Entropy gradient produced by a detached shock in the vicinity of the
axis of symmetry \label{Aaxis}} 

Let $(x_{\rm sh},y_{\rm sh})$ be local coordinates in a  frame chosen in
a way such that $x_{\rm sh}$ is perpendicular to the shock surface, and
$y_{\rm sh}$ is parallel to it. Let us call $\M_1$ the Mach number relative
to the component of the velocity, ahead of and perpendicular to the shock
surface. In our local system of coordinates, 
\begin{equation}
\M_1={v_{x1}\over c_1}\;.
\end{equation}
We shall use the classical convention of using the index ``1" for
quantities ahead of the shock, and the index ``2" for quantities
just after the shock.\\
We write the entropy gradient after the shock as:
\begin{equation}
\nabla S={v_2\over v_{x2}}{\p S_2\over\p y}={v_2\over v_{x2}}
{\p S_2\over\p \M_1}{\p\M_1\over\p y} \;. 
\end{equation}
We can use the classical jump conditions~(\ref{entroRH}) to express the
entropy gradient as a function of $\M_1$ only: 
\begin{eqnarray}
\nabla S&=&{v_2\over v_{x2}}{2\eta\over\gamma-1} {\p
\log \M_1\over\p y}\;,\; {\rm where}
\label{ds2}\\
\eta(\M_1)&\equiv &{2\gamma(\gamma-1)(\M_1^2-1)^2\over
(2+(\gamma-1)\M_1^2)(2\gamma \M_1^2-(\gamma-1))} <1
\end{eqnarray}

We obtain $v_{x1}$ by the vector product of the velocity and the
vector tangent to the shock surface. Let the supersonic flow before the shock 
be described in  cylindrical coordinates by the flow velocity
$v_r(\br,\theta),v_\theta(\br,\theta)$. For the sake of clarity, we denote
the velocity along the shock surface $r_{\rm sh}(\theta)$ by 
$v(\theta)\equiv v(\brs(\theta),\theta)$ and a dot stands for the derivative
with respect to the angle $\theta$: 
\begin{equation}
v_{x1}(\theta)\equiv 
{\brs(\theta) v_r(\theta)-
{\dot\brs}(\theta) v_\theta(\theta)
\over
\left(\brs^2(\theta)+{\dot\brs}^2(\theta)\right)^{1\over2}}\;.
\end{equation}
We now prove that a regular axisymmetric shock surface cannot produce a
local region of uniform entropy behind it, and more precisely, that the
gradient of entropy created by the shock scales like the square of the
azimuthal angle $(\theta-\pi)$ in the vicinity of the axis, for any shape of
the shock which is regular and symmetric. \\
When not specified, the quantities are taken at $\theta=\pi$ in what
follows. We also define $v_{\rm sh}\equiv v_r(\pi)$ and $\br\equiv
\brs(\pi)$. The symmetry condition imposes:
\begin{equation}
{\dot \brs}(\pi)=0
\end{equation}
Using the symmetry of the problem, we make the following expansions
for small values of $(\theta-\pi)$:
\begin{eqnarray}
v_r(\theta)&=&v_{\rm sh}+{(\theta-\pi)^2\over 2}
\left({\ddot\brs}{\p v_r\over\p\br}+{\p^2v_r\over\p\theta^2}\right)\nonumber
\\
& &+{\cal O}\left((\theta-\pi)^4\right)\;,\label{expvr}\\
v_\theta(\theta)&=&(\theta-\pi){\p v_\theta\over\p\theta}+ 
{\cal O}\left((\theta-\pi)^3\right)\;,\label{expvt}\\
\brs(\theta)&=&\brs+{(\theta-\pi)^2\over2}{\ddot \brs}+
{\cal O}\left((\theta-\pi)^4\right)\;.\label{exprs}
\end{eqnarray}

Using the expansions (\ref{expvr}), (\ref{expvt}), and (\ref{exprs}), we
obtain:
\begin{eqnarray}
& &{v_{x1}(\theta)\over v_{\rm sh}}=1-\nonumber\\
&&{(\theta-\pi)^2\over2}
\left\lbrack
\left({{\ddot\brs}\over\brs}\right)^2 + {{\ddot\brs}\over\brs}
\left({2\over v_{\rm sh}}{\p v_\theta\over\p\theta}-
{\brs\over v_{\rm sh}}{\p v_r\over\p\br}\right)
-{1\over v_{\rm sh}}{\p^2v_r\over\p\theta^2}\right\rbrack\nonumber\\
&&+{\cal O}\left((\theta-\pi)^4\right)
\;.\label{vnbrut}
\end{eqnarray}
In order to evaluate this expression, we use the approximation of the
supersonic flow $v_r(\br,\theta),v_\theta(\br,\theta)$ by hyperbolic
trajectories. Derivating the Bernoulli Eq.~(\ref{Bernoulli}) with respect to
$\br$ and $\theta$, we obtain for $\theta=\pi$:  
\begin{eqnarray}
{\p^2 v_r\over\p\theta^2}&=&-{1\over v_{\rm sh}}
\left({\p v_\theta\over\p\theta}\right)^2\;,\\
{\p v_r\over\p\br}&=&-{v_\infty^2\over2v_{\rm sh}\brs^2}\;.
\end{eqnarray}
Replacing these expressions, together with
Eqs.~(\ref{enervr}) and (\ref{dvtheta}) into Eq~(\ref{vnbrut}), the effective
Mach number is: 
\begin{eqnarray}
{\M_{\rm 1}(\theta)\over\M_{\rm 1}(\pi)}=
1&-&{(\theta-\pi)^2\over 2}
\left\lbrack 
{3+4\brs+4\brs^{1\over2}(1+\brs)^{1\over2}\over16\brs(1+\brs)^2}
+ C^2\right\rbrack\nonumber\\
&+&{\cal O}\left((\theta-\pi)^4\right) \;,
\end{eqnarray}
where the positive contribution $C^2$ depends on the second derivative
${\ddot \brs}(\pi)$ of the shock surface on the symmetry axis, and is
defined as: 
\begin{equation}
C\equiv {{\ddot \brs}\over\brs}+{1\over v_{\rm sh}}\left(
{\p v_\theta\over\p\theta} - {\brs\over2}{\p v_r\over \p \br}\right)
\end{equation}
We conclude that {\it for any shape of the detached shock surface}, the
effective Mach number near the symmetry axis decreases at least as the square
of the azimuthal angle $(\theta-\pi)^2$.\\
This result is particularly useful in order to prove that the mass accretion
rate in the subsonic region of Type SF flows decreases to zero with the size
of the accretor.

\begin{figure*}
\epsfxsize=18cm \epsfclipon \epsffile{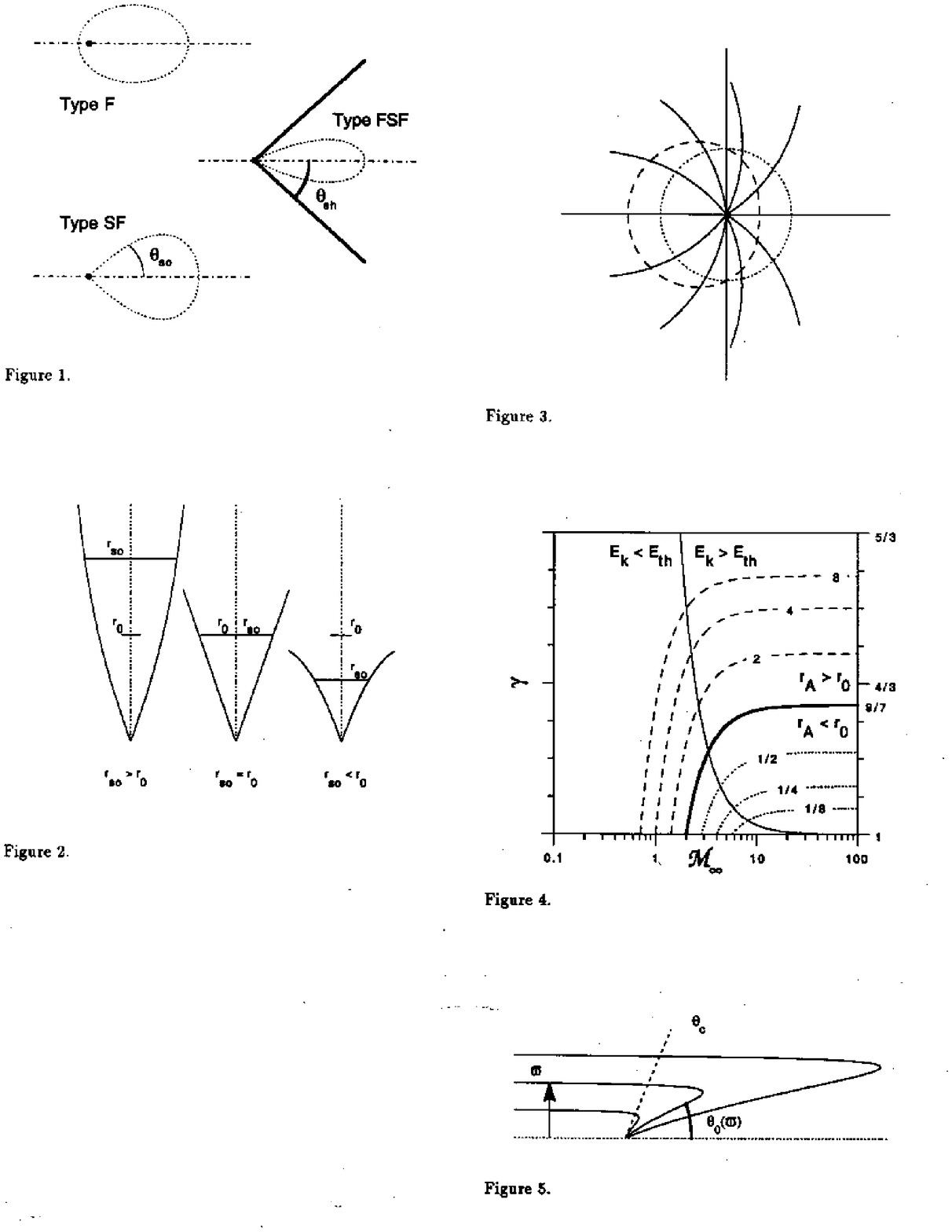}
\end{figure*}

\begin{figure*}
\epsfxsize=18cm \epsfclipon \epsffile{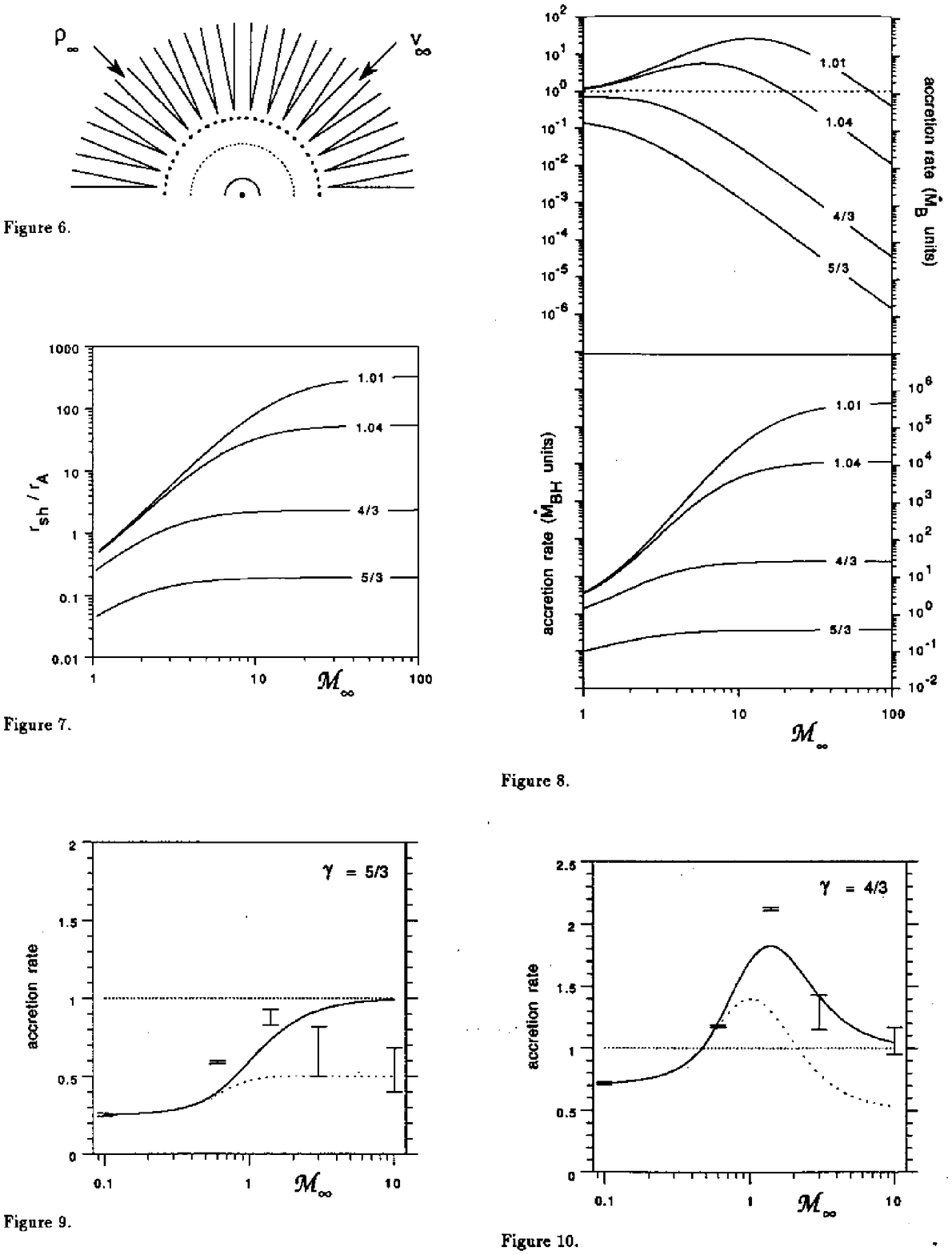}
\end{figure*}

\end{document}